\documentclass[12pt,a4paper,fleqn]{article}
\usepackage[textheight=23cm,textwidth=16cm]{geometry}
\usepackage{amsmath}
\usepackage{amssymb}
\usepackage{graphicx}
\usepackage{cite}
\usepackage[OT4]{polski}
\usepackage[american]{babel}

\begin{document}

\begin{center}

{\LARGE\bf
 Inverse Quantum Chemistry: Concepts and Strategies for Rational Compound Design
}

\vspace{2cm}

{\large
Thomas Weymuth
and Markus Reiher\footnote{corresponding author; e-mail: markus.reiher@phys.chem.ethz.ch}
}\\[4ex]

ETH Z\"urich, Laboratorium f\"ur Physikalische Chemie, \\
Vladimir-Prelog-Weg 2, 8093 Z\"urich, Switzerland

\vspace{1cm}
\end{center}

\begin{abstract}

The rational design of molecules and materials is becoming more and more important. With the advent of powerful computer 
systems and sophisticated algorithms, quantum chemistry plays a decisive role in the design process. While traditional quantum 
chemical approaches predict the properties of a predefined molecular structure, the goal of inverse quantum chemistry is to 
find a structure featuring one or more desired properties. Herein, we review inverse quantum chemical approaches proposed 
so far and discuss their advantages as well as their weaknesses.

\end{abstract}

\vfil
\begin{tabbing}
Date:   \quad \= March 9, 2014 \\
Status:       \> printed in {\it Int.~J.~Quantum Chem.} {\it 114} {\bf 2014} 823--837\\
\end{tabbing}

%
%
%

\newpage

\renewcommand{\baselinestretch}{1.5}
\normalsize

\newpage
\section{Introduction}
\label{sec:intro}

The last decades have witnessed the fast-paced development of a wide range of quantum chemical methods\cite{dyks05}. 
Examples are highly accurate but also computationally demanding wave-function-based approaches like coupled cluster and 
configuration interaction\cite{helg00} and density functional theory\cite{parr89}. The latter allows for the description of 
large molecular systems consisting of hundreds of atoms\,---\,usually at reduced accuracy, however. Today, these computational 
approaches are firmly established in chemistry, and they are crucial for basically all areas of chemical research.

Underlying all these methods is the time-independent Schr\"odinger equation,
\begin{equation}
 \label{eq:schrodinger_equation}
 \hat{H}\Psi = E\Psi,
\end{equation}
where $\Psi$ is the sought-for wave function characterizing the system under investigation, $E$ the energy associated with 
this wave function, and $\hat{H}$ is the Hamiltonian operator. In nonrelativistic quantum chemistry, the Hamiltonian of an 
assembly of (point-like) atomic nuclei and electrons is given by
\begin{equation}
 \label{eq:nonrelativistic_hamiltonian}
 \hat{H} = -\frac{1}{2}\sum_I\frac{1}{M_I}\Delta_I - \frac{1}{2}\sum_i\Delta_i - \sum_{i,I}\frac{Z_I}{r_{iI}} + \sum_{i,j>i}\frac{1}{r_{ij}} + \sum_{I,J>I}\frac{Z_IZ_J}{r_{IJ}}
\end{equation}
(in Hartree atomic units, in which the elementary charge, the mass of the electron, the reduced Planck constant, and 
$4\pi\varepsilon_0$ take a value of one\,---\,used throughout this work, unless otherwise stated). The indices $I$, $J$ and 
$i$, $j$ run over all nuclei and electrons, respectively. $M_I$ denotes the mass of nucleus $I$, $Z_I$ its charge 
number, and $r_{ij}$  is the spatial distance between particles $i$ and $j$. The Laplacian is defined as
\begin{equation}
 \Delta_i = \frac{\partial^2}{\partial x_i^2} + \frac{\partial^2}{\partial y_i^2} + \frac{\partial^2}{\partial z_i^2}.
\end{equation}
The first two terms in Eq.~(\ref{eq:nonrelativistic_hamiltonian}) are the kinetic energy operators of the nuclei and electrons, 
while the last three terms describe the electron--nucleus attraction, the electron--electron repulsion, and the nucleus--nucleus 
repulsion, respectively. The rather complicated Hamiltonian of Eq.~(\ref{eq:nonrelativistic_hamiltonian}) is usually simplified 
by invoking the well-known Born--Oppenheimer approximation, in which the nuclear coordinates are treated as (fixed) parameters. 
We may then define the electronic Hamiltonian
\begin{equation}
 \label{eq:nonrelativistic_electronic_hamiltonian}
 \hat{H}_{\rm el} = - \frac{1}{2}\sum_i\Delta_i - \sum_{i,I}\frac{Z_I}{r_{iI}} + \sum_{i,j>i}\frac{1}{r_{ij}} + \sum_{I,J>I}\frac{Z_IZ_J}{r_{IJ}},
\end{equation}
and the corresponding electronic Schr\"odinger (eigenvalue) equation
\begin{equation}
\label{elS-Gl}
 \hat{H}_{\rm el}\Psi_{\rm el} = E_{\rm el}\Psi_{\rm el}.
\end{equation}
The molecular structure, defined as an assembly of atomic nuclei fixed in space, is a direct consequence of this approximation.

For a given assembly of atomic nuclei and electrons, the nonrelativistic electronic Hamiltonian is unequivocally defined. The 
methods mentioned in the beginning aim at an approximate solution of the electronic Schr\"odinger equation. In principle, all 
observables and molecular properties of interest can then be calculated from the wave function, which contains all information 
that can be known about the system\cite{cohe09}. Therefore, quantum chemical methods provide us with a wealth of information 
about a certain system, but only \textit{after} this very system has been specified in terms of the nuclear framework. For 
the design of new functional molecules or materials one encounters the reverse situation, where a desired property is known 
and a molecule featuring this very property is searched for. Given the ``forward'' direction from structure to property, 
extensive screening of structures is necessary to find a molecule with a predefined property (one example for this approach 
is drug design). It is thus highly desirable to develop computational approaches which would render such time- and cost-intensive 
trial-and-error procedures obsolete. In order to achieve this goal, the direction needs to be reversed, i.e., from property 
to structure. Such approaches are called ``inverse approaches''.

Inverse problems do not only occur in the field of chemistry, but play an important role in diverse areas such as geophysics 
(where the composition of the Earth's mantle is infered from gravimetric measurements\cite{geophys_j_int_2007_171_1118}), 
medical imaging (where, for example, attempts are made to clarify the internal structure of tissues by ultrasonic pressure 
waves\cite{engl97}) and computer vision (where three-dimensional data from lower-dimensional images is constructed\cite{computer-aided_design_2005_37_509}). 
Accordingly, the theory of inverse problems constitutes a whole branch of mathematics with important contributions made in 
the first half of the 20th century by Russian researchers (see Ref.~\cite{chad77} for an extensive bibliography). Clearly, 
since inverse problems represent an important aspect in many fields of research, the study of inverse problems in quantum 
chemistry may well benefit from developments made in other fields\cite{int_j_quantum_chem_2009_109_2456}.

As we will see, the greatest challenge an inverse chemical application faces is the huge size of chemical space, i.e., the 
space of molecules accessible by contemporary synthetic protocols. It has been estimated that this size is between 10\textsuperscript{20} 
and 10\textsuperscript{24} molecules\cite{j_chem_inf_comput_sci_2003_43_374}. However, an estimation of the total size of 
chemical space highly depends on the assumptions made; for example, the number of molecules having up to 30 carbon, nitrogen, 
oxygen, and sulfur atoms has been estimated to be larger than 10$^{60}$\cite{med_res_rev_1996_16_3}, while the number of 
proteins that could theoretically exist is roughly 10$^{390}$\cite{nature_2004_432_824} (for an average size of 300 amino 
acid residues per protein). This might be the reason for the fact that inverse quantum chemical approaches have only emerged 
during the past 20 years, although the basic idea has been presented much earlier\cite{chem_soc_rev_1973_2_21}. It is the 
purpose of this work to review all existing approaches in this field and to analyze their potential for future applications.

We will first review in section \ref{sec:ist} the mathematical concepts developed for inverse Sturm--Liouville problems, 
where important results emerged as early as 1929 and strongly influenced research carried out between 1950 and 1970 on 
inverse quantum scattering. Then, we review inverse techniques which have either already found widespread application in rational 
compound design or exhibit a great potential.
We discuss the inverse perturbation analysis in section \ref{sec:ipa} and the use of model equations in section \ref{sec:modeq}. 
Next, we move on to more recent approaches, which rely on sophisticated sampling and optimization techniques, 
such as the optimization of wave functions in section \ref{sec:optimwf}, 
quantitative structure--activity relationships in section \ref{sec:qsar}, 
and the optimization of solid state compounds in 
sections \ref{sec:ssc} and \ref{sec:ibsa}. We review two very recent approaches, namely the linear combination of atom-centered 
potentials in section \ref{sec:lcacp} and alchemical potentials in section \ref{sec:alchpot}. Finally, we also highlight 
developments from our research group in sections \ref{sec:tracking} and \ref{GdMC-sec} followed by general conclusions in section \ref{sec:conclusion}.

\section{Inverse Spectral Theory}
\label{sec:ist}

Inverse spectral theory is the branch of mathematics which studies what can be deduced about the structure of (differential) 
operators given that some of their properties are known\cite{trub87}. In quantum chemistry, this translates to the question whether 
it is possible to reconstruct the Hamiltonian from a predefined set of eigenvalues.

The Soviet--Armenian physicist Viktor Ambarzumian\cite{name} investigated to which extent a differential equation is defined 
by its eigenvalue spectrum\cite{zeitschrift_f_phys_1929_53_690}. Basic inverse questions\,---\,although not explicitly stated 
as such\,---\,have been studied earlier in a purely mathematical context (see, for example, Ref.~\cite{proc_cam_phil_soc_1926_23_44}). 
In fact, Lord Rayleigh briefly discussed an inverse problem already in 1877 in his work about acoustic waves\cite{rayl45}. 
Moreover, Ambarzumian was well aware of the fact that the ability of reconstructing the Hamiltonian from its eigenvalue spectrum 
would yield great insight into the structure and properties of matter\cite{zeitschrift_f_phys_1929_53_690}. He studied the 
simple case of an oscillating string, mathematically modeled by a differential equation of second order,
\begin{equation}\label{eq:ambarzumian}
 -\frac{{\rm d}^2\phi(x)}{{\rm d}x^2} + q(x)\phi(x) = \alpha\phi(x); \hspace{1cm} 0\leq x \leq 1
\end{equation}
with the boundary conditions
\begin{equation}
 \frac{{\rm d}\phi(0)}{{\rm d}x} = \frac{{\rm d}\phi(1)}{{\rm d}x} = 0,
\end{equation}
where $\phi(x)$ describes the oscillation as a function of position $x$. $q(x)$ is a continuous function, and $\alpha$ is a 
constant. He showed that the series of eigenvalues
\begin{equation}
 \alpha_n = n^2\pi^2; \hspace{1cm} n \in \mathbb{N}
\end{equation}
is only consistent with the differential equation, Eq.~(\ref{eq:ambarzumian}), if $q(x) \equiv 0$. It is easily seen that 
Eq.~(\ref{eq:ambarzumian}) resembles the Schr\"odinger equation for a particle in a one-dimensional box if we take $\alpha$ 
as proportional to the energy eigenvalue and $q(x)$ as a position-dependent potential. Note, however, that the boundary 
conditions employed by Ambarzumian are different from the ones of the particle in a one-dimensional box. Ambarzumian studied 
so-called natural boundary conditions, where the first derivative of the function must vanish at the box boundaries, whereas 
for the 'particle in the box' model the function itself must be zero at the boundaries.

\begin{figure}
  \begin{center}
    \includegraphics[scale=0.2]{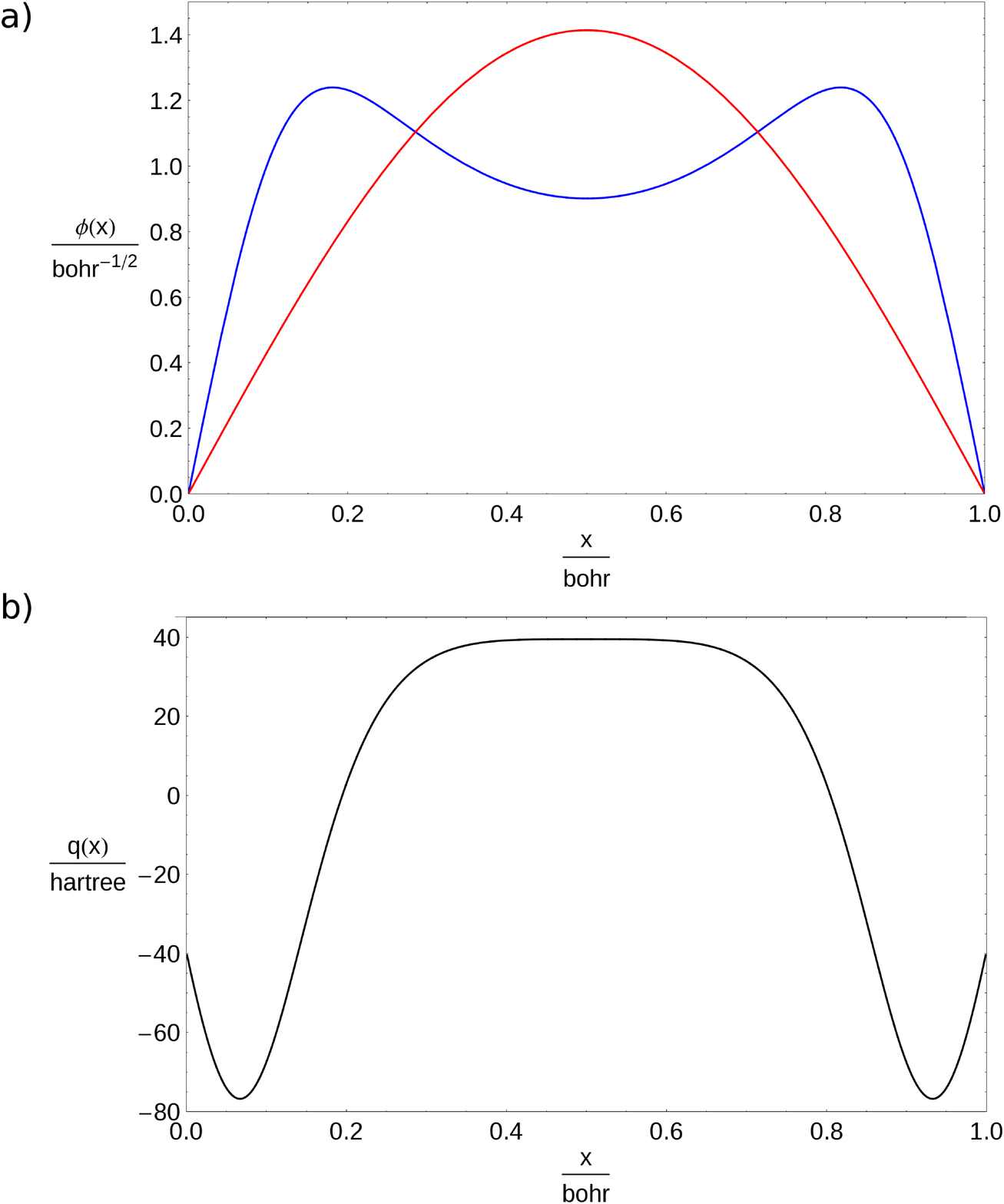}
  \end{center}
  \caption{a) Normalized wave functions for an electron in a one-dimensional box of length 1\,bohr. Red: ground state wave 
function when no potential energy term is present in the Hamiltonian. Blue: ground state wave function if an additional potential 
is introduced such that the lowest eigenvalue is shifted from $\pi^2$ to $3\pi^2$ while all others remain unchanged. b)
Potential $q(x)$ producing the wave function of Eq.~(\ref{eq:trubowitz}) with $\mu^2 = 3$.\label{fig:trub}}
\end{figure}

From the results of Ambarzumian, it follows that we can reconstruct the potential $q(x)$\,---\,and concomitantly the full 
Hamiltonian\,---\,if we know the (complete) set of energy eigenvalues (see also Ref.~\cite{trub87}). For example, we can 
choose a potential $q$ such that the resulting Hamiltonian features an eigenvalue spectrum of $\mu^2\pi^2, n^2\pi^2$ with 
$n \in \mathbb{N}$, $1 < \mu < 2$ and $n>\mu$ (i.e., $n=1$ is not possible and the lowest eigenvalue is shifted from $n=1$ to $n'=\mu$). 
Thus, we can selectively shift only the lowest eigenvalue while leaving all others 
unchanged. In this case, the (unnormalized) wave function corresponding to the lowest eigenvalue $\mu^2\pi^2$ can be 
shown\cite{trub87} to be given by
\begin{equation}\label{eq:trubowitz}
 \phi_\mu(x) = \frac{{\rm sin}(\pi x)}{\displaystyle \left[ \frac{{\rm sin}(\mu(1-x)) - {\rm sin}(\mu x)}{{\rm sin}(\mu)}, {\rm sin}(\pi x) \right]},
\end{equation}
where the denominator is defined as
\begin{equation}
 \left[ f(x), g(x) \right] = f(x)\frac{{\rm d}g(x)}{{\rm d}x} - \frac{{\rm d}f(x)}{{\rm d}x}g(x),
\end{equation}
the so-called Wronskian of $f$ and $g$. The function given in Eq.~(\ref{eq:trubowitz}) is depicted in Fig.~{\ref{fig:trub}}\,a) 
for $\mu^2 = 3$, and compared to the (regular) ground state wave function of an electron in a one-dimensional box, i.e., for $q'(x) = 0$.

The corresponding potential can easily be constructed numerically by inserting Eq.~(\ref{eq:trubowitz}) into the Schr\"odinger 
equation for a particle in a one-dimensional box and solving for $q$. It is plotted in Fig.~{\ref{fig:trub}}\,b).

Several different methods to construct a Hamiltonian from a given energy spectrum have been developed. However, in 1946, 
the Swedish mathematician G\"oran Borg generalized Ambarzumian's work\cite{acta_mathematica_1946_78_1}. In his seminal paper, 
Borg showed that in the general case, one {\it cannot} determine the function $q(x)$ from only one eigenvalue spectrum. 
Instead, one also needs the spectrum of a ``complementary'' eigenvalue problem (e.g., the same differential equation with 
different boundary conditions) in order to fully determine the corresponding differential operator. Thus, in the general 
case we cannot deduce the form of the Hamiltonian by predefining a suitable set of energy eigenvalues, as is done in the 
above example. It was pointed out that such an inversion is in general not possible, since different Hamiltonians (representing 
different molecules) can actually feature comparable properties\cite{nat_mater_2004_3_838,phys_rev_lett_2005_95_153002}. 
Even if it would be possible to strictly invert the Schr\"odinger equation in the most general way, it can well be imagined 
that among the infinitely many different molecules, there will always be molecules that feature almost identical values for 
a given property. One can thus intuitively understand that for a given application, several molecules can exhibit the correct 
property. Therefore, different approaches have to be devised to tackle this difficult problem. In the following, we will 
review the most important of these approaches.

\section{Inverse Perturbation Analysis}
\label{sec:ipa}

In molecular spectroscopy, one is often interested in determining the potential leading to a certain energy spectrum as 
accurately as possible.
Nowadays, there exist many different approaches to accomplish this task\cite{j_mol_spectrosc_1973_46_25,j_mol_spectrosc_1974_51_321,
j_chem_phys_1995_102_2282, j_chem_phys_1995_103_9713}, and there are also specialized computer programs available (see, 
e.g., Ref.~\cite{comp_phys_comm_2000_128_622}). For the sake of brevity, we shall not extensively dwell here on the details 
of all these approaches. However, we would like to highlight one method, which is very general in its scope of use and can 
also serve as motivation for the development of more advanced inverse approaches, as we shall see.

In 1975, Kosman and Hinze developed a method they called Inverse Perturbation Analysis (IPA) which allows one to calculate 
accurate potential energy curves\cite{j_mol_spectrosc_1975_56_93}. They started from the one-dimensional Schr\"odinger equation 
for a diatomic molecule (note, however, that the method itself is not limited to diatomics),
\begin{equation}\label{eq:radialeq}
 \left( -\frac{1}{2\mu}\frac{{\rm d}^2}{{\rm d}r^2} + U(r) \right) R(r) = E R(r),
\end{equation}
where $\mu$ is the reduced mass of the molecule under consideration, $r$ the internuclear distance, $R(r)$ the nuclear wave 
function, and $E$ the corresponding energy eigenvalue. $U(r)$ is the potential energy function, which is usually not known 
exactly. However, many approximate potentials are known, which are often very close to the unknown exact potential like the 
well-known Rydberg--Klein--Rees potential. One of the central ideas of IPA is now to write the exact potential as a sum of a 
known approximate potential $U^{(0)}$ and a (small) correction term $\Delta U$,
\begin{equation}\label{eq:ipapot}
 U(r) = U^{(0)}(r) + \Delta U(r).
\end{equation}
If the approximate potential is a good approximation to the exact curve, the correction term is expected to be small, and can 
consequently be searched for with an iterative perturbative approach, where $\Delta U$ is the perturbation. First, 
Eq.~(\ref{eq:radialeq}) is solved numerically for the wave functions $R^{(0)}$ and energy eigenvalues $E^{(0)}$ with the 
approximate potential $U^{(0)}$. Since the true energy eigenvalues $E$ can be obtained experimentally by spectroscopic 
techniques, one can calculate an energy correction
\begin{equation}
 \Delta E = E - E^{(0)},
\end{equation}
which can be approximated by the first-order energy correction
\begin{equation}\label{eq:ipade}
 \Delta E = \left\langle R^{(0)}(r)\left|\Delta U(r)\right|R^{(0)}(r)\right\rangle.
\end{equation}
If one expands the potential correction into a set of predefined suitable basis functions, $f_i(r)$,
\begin{equation}
 \Delta U(r) = \sum_i c_i f_i(r),
\end{equation}
then the energy correction becomes
\begin{equation}
 \Delta E = \sum_i c_i \left\langle R^{(0)}(r)\left|f_i(r)\right|R^{(0)}(r)\right\rangle,
\end{equation}
which can be solved by standard methods to obtain the best set of coefficients $c_i$. Once this set is obtained, one can use 
the coefficients to improve on the representation of the actual potential energy curve by employing Eq.~(\ref{eq:ipapot}). 
With this improved potential, one can iterate the procedure outlined above to obtain an even better potential function. Thus, 
one can stepwise improve the description of the potential energy until the energy correction term $\Delta E$ is below a 
given threshold.

In their original paper, Kosman and Hinze successfully applied their method to HgH. However, they found that their results 
heavily depend on the particular choice of basis functions $f_i(r)$. While Kosman and Hinze employed global polynomials (i.e., 
polynomials defined over the entire definition range of $r$, $-\infty\leq r \leq \infty$), Hamilton and coworkers relied on 
local Gaussian functions and found accurate and rapidly convergent results several years later\cite{j_chem_phys_1986_85_5151}. 
Still, Hamilton and coworkers experienced some numerical instabilities (the results were very sensitive to changes in the 
parameters of the basis functions). These instabilities are a characteristic of inverse problems: they are typically ill-conditioned. 
In our case, the behavior of the potential energy function $U(r)$ in regions where the wave function $R(r)$ is zero is not 
determined, as can be seen by inspection of Eq.~(\ref{eq:ipade}), and, therefore, the solution of this equation is not unique. 
We may recall that it can be mathematically proven that a single energy eigenvalue spectrum is in general not sufficient to 
determine the underlying potential uniquely. To remedy this problem of numerical instability, Wu and Zhang proposed to solve 
Eq.~(\ref{eq:ipade}) by means of a singular value decomposition and found that their approach is fast, accurate and numerically 
stable\cite{chem_phys_lett_1996_252_195}.

We thus have a working method to produce accurate molecular potential energy functions, given the exact energy spectrum is 
known and a good approximate potential can be found as a starting point. This is certainly very interesting for the inverse 
point of view. However, we are still confronted with two major problems: first, we are usually not that much interested in a 
molecule featuring a given energy eigenvalue spectrum, but rather in a molecule with a specific property (such as a specific
dipole moment). These properties can, however, usually not be directly related to a specific energy spectrum. Furthermore, 
even if we could somehow overcome this problem and come up with a certain molecular potential, it would not automatically be 
clear how the underlying molecular structure would be composed of atomic nuclei and electrons. However, this problem could 
be solved in a quite elegant fashion by employing some sort of atom-based basis functions, such that each basis function would 
represent one (abstract) atom\,---\,the actual type of the atom would then be determined by the final expansion coefficient 
$c_i$. In the above optimization procedure, one would, of course, not only have to search for the expansion coefficients, but 
also for the position at which the basis functions are centered (which enter as a parameter in these local functions and denote 
the position of the nuclei in space), unless one uses a comparatively large number of such basis functions, such that they can 
simply be distributed uniformly in space. At a position where no nuclei should be, we can expect that the respective expansion 
coefficient is close to zero. In fact, such approaches have been emerging during the past few years, see sections \ref{sec:lcacp} 
and \ref{sec:alchpot}.

Even though there appears to be no general solution of the first problem mentioned above, we shall see in the sections to come 
that we might find other ways of directly relating the molecular potential to properties other than the energy eigenvalue spectrum, 
thereby making this approach a very promising one. We close this section with a reference to work by Politzer and Murray \cite{poli02,poli11}
who elaborated on the central role of the electrostatic potential created by atomic nuclei and electrons as the determinant
of molecular properties and reactive behavior.

\section{Model Equations}
\label{sec:modeq}

A completely different and very interesting approach is the development of (simple) model equations. This technique has been 
successfully employed by Marder \textit{et al.}~in 1991. They describe possibilities to increase the first electronic 
hyperpolarizability of conjugated organic molecules\cite{science_1991_252_103}. We illustrate the underlying design approach 
in the following, but note that the design of molecules with optimal optoelectronic properties has been studied by many groups\cite{j_phys_condens_matter_200_15_R897,mol_phys_2003_101_2279,phys_rev_lett_1999_82_3304}.

If a molecule is placed in an external electric field, its charge density is polarized. When the external electric field has 
only a comparatively small effect on the electronic structure of a given molecule, the polarization $\boldsymbol{P}$ can 
be written as a power series
\begin{equation}
 \boldsymbol{P} = \alpha\boldsymbol{E} + \beta\boldsymbol{E}^2 + \gamma\boldsymbol{E}^3 + \mathcal O(\boldsymbol{E}^4),
\end{equation}
where $\boldsymbol{E}$ is the external electric field, and the constants $\alpha$, $\beta$, and $\gamma$ are the polarizability, 
first hyperpolarizability, and second hyperpolarizability, respectively. We note in passing that in the general case, the 
(hyper-)polarizabilities are not scalar quantities but instead described by tensors. However, for the following discussion, our 
approximation does not constitute any loss of generality. Since the electric field constitutes only a minor perturbation of the 
molecule, perturbation theory is a suitable method to calculate $\alpha$, $\beta$, and higher order corrections (in cases 
where the wave length of irradiation is much larger than the size of the molecule, the perturbation Hamiltonian describes the interaction of 
the molecular dipole moment with the external electric field). The first hyperpolarizability $\beta$ can thus be obtained by 
correcting the wave function to second order in the electric field\cite{bera91}. The resulting expression is rather complicated, 
involving sums over all occupied orbitals. However, it is often observed that the first hyperpolarizability is dominated by 
contributions from only the first few excited states. This motivates a simplification by taking only the ground state ``g'' 
and the first excited state ``e'' into account. The resulting two-state approximation reads\cite{bera91},
\begin{equation}\label{eq:beta}
 \beta \propto \frac{X_{\rm ge}^2 \left( X_{\rm gg} - X_{\rm ee} \right)}{\left( E_{\rm g} - E_{\rm e} \right)^2},
\end{equation}
with the matrix elements defined as
\begin{equation}
 X_{ij} = \left\langle \Psi_i \right| \boldsymbol{r} \left| \Psi_j \right\rangle; \hspace{1cm} i,j \in \{{\rm e, g}\}.
\end{equation}
The influence of the three quantities $X_{\rm ge}^2$, $X_{\rm gg} - X_{\rm ee}$, and $1/\left( E_{\rm g} - E_{\rm e} \right)^2$ 
on the first hyperpolarizability can be studied at a simple two-orbital system --- following Ref.\ \cite{bera91}, for which we may 
write the wave function as a linear combination of these two orbitals,
\begin{equation}
 \Psi = c_1\phi_1 + c_2\phi_2.
\end{equation}
This linear combination is to be understood as a quantum-mechanical model wave function constructed from orbitals $\phi_i$ 
defined on different parts of the complete molecule (this model is in the spirit of classical molecular orbital theory like H\"uckel theory,
i.e., a proper construction of a wave function as an antisymmetrized Hartree product and the consideration of spin are
left aside).

\begin{figure}
  \begin{center}
    \includegraphics[scale=0.2]{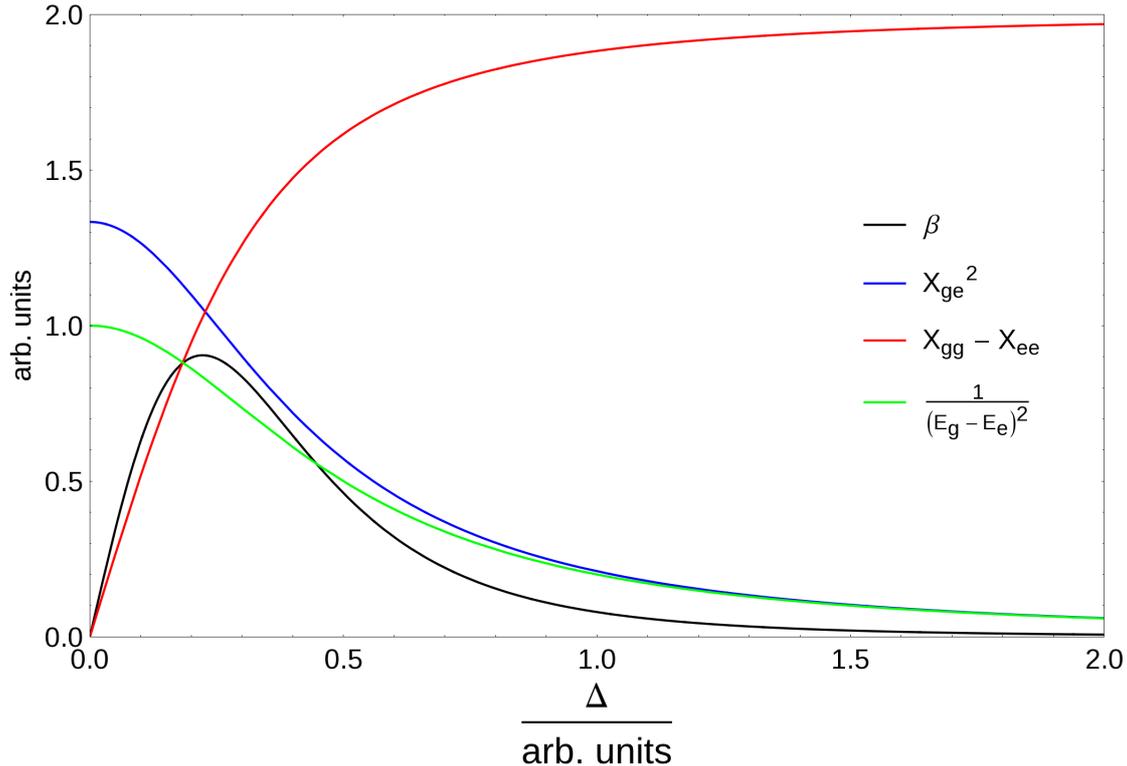}
  \end{center}
  \caption{The first hyperpolarizability $\beta$ of a simple two-orbital system --- calculated from its components 
$X_{\rm ge}^2$, $X_{\rm gg} - X_{\rm ee}$, and $1/\left( E_{\rm g} - E_{\rm e} \right)^2$ according to Eq.~(\ref{eq:beta}) ---
as a function of $\Delta$, where $2\Delta$ is the energy difference between the two orbital energies. In this figure, the 
overlap between the two orbitals as well as the coupling strength $t$ were set to 0.5.\label{fig:modeleq}}
\end{figure}

We now define the energy expectation values
\begin{equation}
 \left\langle \phi_1 \right| H \left| \phi_1 \right\rangle = \Delta,
\end{equation}
and
\begin{equation}
 \left\langle \phi_2 \right| H \left| \phi_2 \right\rangle = -\Delta,
\end{equation}
such that the energy difference between the two orbitals $\phi_1$ and $\phi_2$ is $2\Delta$. The coupling strength between 
the two orbitals is then defined as
\begin{equation}
 \left\langle \phi_1 \right| H \left| \phi_2 \right\rangle = t.
\end{equation}
With these definitions, we can write the time-independent Schr\"odinger equation in matrix form as
\begin{equation}\label{eq:hmat}
\begin{pmatrix}
  \Delta - E & t          \\
  t          & -\Delta -E 
\end{pmatrix}
\begin{pmatrix}
  c_1 \\
  c_2  
\end{pmatrix} =
\begin{pmatrix}
  0 \\
  0  
\end{pmatrix}.
\end{equation}
For this system of homogeneous equations to have a nontrivial solution, i.e., $c_1$, $c_2 \neq 0$, the determinant of the 
2$\times$2 matrix must be zero, which yields an equation to determine the energy $E$. With the two solutions
\begin{equation}
 E_{\pm} = \pm \sqrt{\Delta^2 + t^2}.
\end{equation}
and the first equality of Eq.~(\ref{eq:hmat}),
\begin{equation}
 \left( \Delta - E_{\pm} \right)c_1^{(\pm)} + tc_2^{(\pm)} = 0,
\end{equation}
we can establish the relation
\begin{equation}
 \frac{c_1^{(\pm)}}{c_2^{(\pm)}} = \frac{E_{\pm} - \Delta}{t} = \pm\sqrt{1 + \left( \frac{\Delta}{t} \right)^2} - \frac{\Delta}{t}.
\end{equation}
For each of the two energy eigenvalues, we get a separate set of coefficients $c_i$, such that we obtain two wave functions,
\begin{equation}
 \Psi_{\rm g} = \Psi_{-} = c_1^{(-)}\phi_1 + c_2^{(-)}\phi_2,
\end{equation}
and
\begin{equation}
 \Psi_{\rm e} = \Psi_{+} = c_1^{(+)}\phi_1 + c_2^{(+)}\phi_2.
\end{equation}
Expressions for the individual coefficients can be obtained from the normalization condition, i.e., 
\begin{equation}
 \left\langle \Psi_{\pm} | \Psi_{\pm} \right\rangle = 1.
\end{equation}
We can thus calculate expressions for $X_{\rm ge}^2$, $X_{\rm gg} - X_{\rm ee}$, and $1/\left( E_{\rm g} - E_{\rm e} \right)^2$ 
as well as for $\beta$ as a function of $\Delta$. These functions are given in Fig.~\ref{fig:modeleq}. We see that the 
combination of the individual expressions leads to a maximum of $\beta$ at about $\Delta = 0.25$ when the overlap between 
$\phi_1$ and $\phi_2$ and the coupling strength $t$ are set to 0.5. The task is now to find a molecule corresponding to these 
values, which is by far not trivial.

In their publication, Marder \textit{et al.}~also exploited the mathematical expression for the first hyperpolarizability $\beta$ 
given in Eq.~(\ref{eq:beta}), but for a somewhat more sophisticated model system consisting of four orbitals (corresponding 
to donor, acceptor, and bridge orbitals)\cite{science_1991_252_103}. Marder {\it et al.}~could relate the above expressions 
to actual molecular building blocks, in such a way that they were able to come up with specific design strategies to construct 
the desired molecules\cite{science_1991_252_103}.

Unfortunately, the deduction of an actual molecular (sub-)structure is not straightforward at all, but requires often a 
considerable amount of chemical knowledge and intuition. Especially this last fact is a notable drawback, limiting the 
applicability of this method to a rather narrow subspace of chemical compound space. Therefore, this approach suffers from 
the same fundamental problem as the QSPR techniques (see below).

\section{Optimized Wave Functions}
\label{sec:optimwf}

\begin{figure}
  \begin{center}
    \includegraphics[scale=0.2]{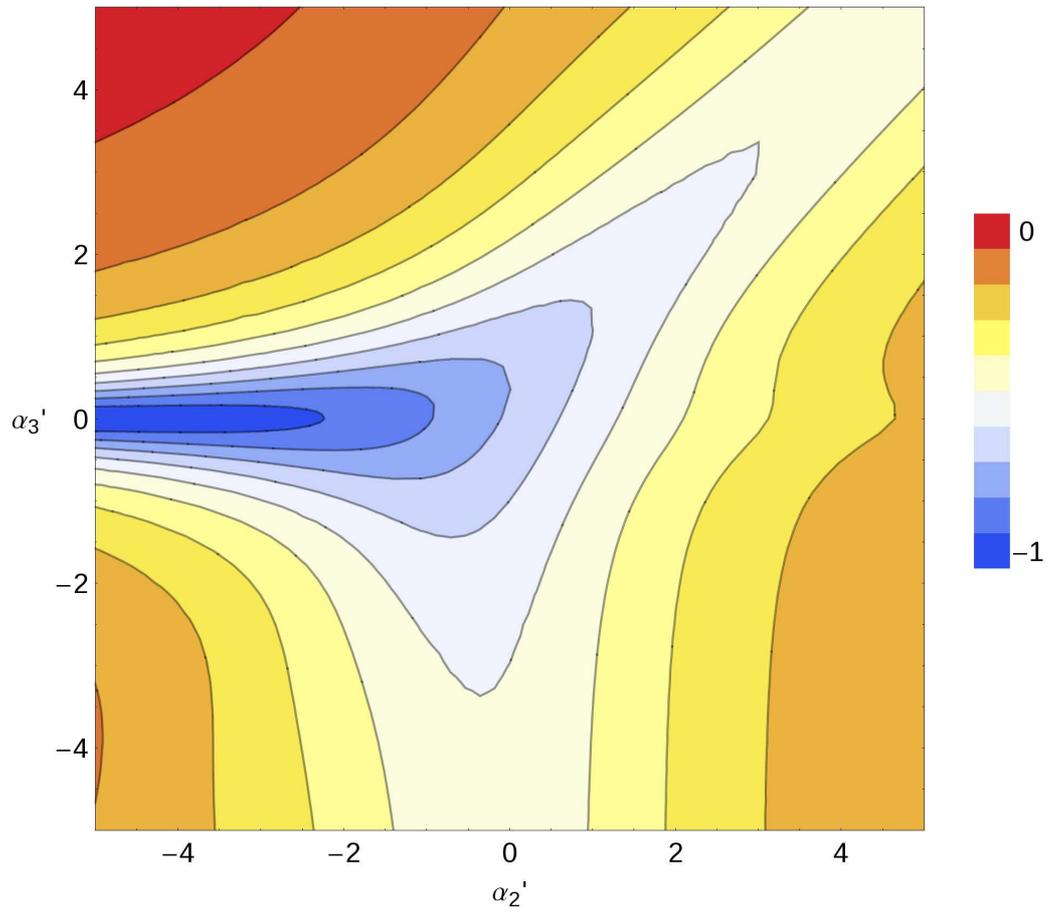}
  \end{center}
  \caption{Dependence of the transition dipole moment $O_{\rm 23}$ on the two parameters $\alpha_2'$ and $\alpha_3'$, respectively. 
All data are given in arbitrary units.\label{fig:optimwf}}
\end{figure}

This severe limitation has long been recognized\cite{j_phys_chem_1996_100_10595}. In 1996, Kuhn and Beratan proposed a new 
strategy in order to overcome this problem, which can be regarded as a first truly inverse approach in rational compound 
design\cite{j_phys_chem_1996_100_10595}. They suggested to search for an optimal molecular structure by optimizing the 
corresponding mathematical objects describing it, i.e., the Hamiltonian and its wave function. The authors also point out that 
this inverse problem is ill-conditioned as not enough information is present in order to carry out the optimization (recall 
Section \ref{sec:ipa}). This problem can be lifted by introducing additional constraints. For example, bonding in molecules 
is local by nature, leading to characteristic patterns in the Hamiltonian matrix. This can be conveniently illustrated by the 
example of the simple H\"uckel molecular orbital method\cite{zeitschr_elektrochem_1937_43_752,zeitschr_elektrochem_1937_43_827}, 
where a linear conjugated hydrocarbon chain corresponds to a tridiagonal Hamiltonian matrix due to a nearest-neighbor approximation. 
The overlap of basis functions located at distant atoms is in fact usually negligibly small, and thus also in more advanced 
electronic structure theories the Hamiltonian matrix is usually sparse. With the H\"uckel method, for a linear chain of three 
orbitals centered at $x_i = \{0, 1, 2\}$, we can write the Hamiltonian in matrix form as
\begin{equation}
 H = \begin{pmatrix}
  \alpha_1 & \beta_1  & 0       \\
  \beta_1  & \alpha_2 & \beta_2 \\
  0        & \beta_2  & \alpha_3
 \end{pmatrix},
\end{equation}
where the $\alpha_i$ represent the orbital energies and the $\beta_i$ the coupling between adjacent orbitals. Since we are 
free in our choice of energy origin, we can arbitrarily set one of the $\alpha_i$ to zero. Furthermore, we can also choose 
any suitable unit for the energy, which allows us to set one of the $\beta_i$ to $-1$. With this, we can write
\begin{equation}
  H \rightarrow H' = \begin{pmatrix}
  0        & -1       & 0       \\
  -1       & \alpha_2'  & \beta_2' \\
  0        & \beta_2'  & \alpha_3'
 \end{pmatrix}.
\end{equation}
The task is now to find values for the remaining parameters such that a given property is optimized. As an example, Kuhn and 
Beratan chose the electric dipole operator, the matrix elements of which can be written as
\begin{equation}
 \mu_{km} = -\sum_{i=1}^3x_ic_i^{(k)}c_i^{(m)},
\end{equation}
if overlap contributions are neglected ($c_i^{(k)}$ is the expansion coefficient of the $i$-th orbital in wave function $k$ 
as in section \ref{sec:modeq}) and optimized the transition dipole moment $\mu_{23} = c_2^{(2)}c_2^{(3)} + 2c_3^{(2)}c_3^{(3)}$. 
Within our simple approach, $\mu_{23}$ is a function of only three variables, namely $\alpha_2'$, $\alpha_3'$, and $\beta_2'$. 
If we arbitrarily set $\beta_2' = \beta_1 = -1$, we can plot $\mu_{23}$ as a function of only $\alpha_2'$ and $\alpha_3'$, 
as shown in Fig.~\ref{fig:optimwf}. It is straightforward to optimize $\mu_{23}$ with respect to these two parameters. For 
$\alpha_2' \geq -5$, one finds that $\mu_{23}$ is minimized for $\alpha_2' = -5.00$ and $\alpha_3' = 0.00$, which yields 
$\mu_{23} = -0.97$\cite{j_phys_chem_1996_100_10595}. The global minimum, however, appears to be at $\alpha_2' = -4848.59$ and 
$\alpha_3' = 0.00$. With these values, one finds $\mu_{23} = -1.00$.

In their pioneering study, Kuhn and Beratan proposed also to optimize not only the elements in the Hamiltonian matrix, but also 
the coefficients of the wave function (while the energy eigenvalues were held fixed). They applied their method to a few very 
simple ``toy'' examples like the one given above in order to demonstrate the general concept.  However, for many years after 
its publication, the method could not be applied to find an actual molecule exhibiting a desired property, as was hoped by 
the authors. The problem was not so much the sheer size of chemical compound space, but rather the fact that it is extremely 
difficult to construct a molecule from its Hamiltonian matrix or wave function\cite{sci_china_ser_b_2009_52_1769}. However, 
in 2006, Beratan and coworkers established a general framework which can circumvent this problem (see Section \ref{sec:lcacp}).

\section{Quantitative Structure--Activity Relationships}
\label{sec:qsar}

Quantitative Structure--Activity Relationships (QSAR) and the closely related Quantitative Structure--Property Relationships 
(QSPR) are central for design attempts in chemistry, biology, and materials research\cite{chem_rev_2012_112_2889,curr_med_chem_2011_18_2517,chem_rev_1996_96_1027,chem_soc_rev_1995_24_279,
excli_journal_2009_8_74,chem_rev_2010_110_5714,chempluschem_2012_77_507}. The origins of QSAR trace back to as early as 
1863, when Cros described a relation between the toxicity of primary aliphatic alcohols and their solubility in water\cite{cros_phd}. 
However, the actual mechanistic basis of contemporary QSAR approaches was laid in 1964 by Hansch and Fujita with the development 
of the linear Hansch equation\cite{j_am_chem_soc_1964_86_1616}, which was based on the famous work of Hammett, who proposed 
an equation to relate equilibrium constants of a particular reaction involving benzene derivatives to some structural parameter, 
which depends on the particular type of benzene substituents, and a reaction parameter, which depends on the type of reaction under study\cite{chem_rev_1935_17_125,j_am_chem_soc_1937_59_96}.

The basic assumption of QSPR is that every physical, chemical, and biological property depends in a systematic way on the 
underlying molecular structure. The functional form of this dependence is then searched for by trying to establish a model 
relating a given property to one or more descriptive parameters of the structure (so called descriptors). In a typical QSPR 
study, a training set of molecular structures and corresponding properties is taken as input. The descriptors are then 
usually calculated directly from the molecular structure. As descriptors, diverse quantities such as the molecular weight, 
the number of ring systems, the dipole moment, or the solvent-accessible molecular surface can be employed\cite{chem_soc_rev_1995_24_279}. 
Even though one could also use descriptors which are obtained experimentally, it is advantageous to employ descriptors which 
can be obtained directly from the (three-dimensional) molecular structure without the need of time-consuming measurements. 
Once all desired descriptors have been obtained, a statistical analysis of these data is carried out in order to find a 
mathematical relation between the descriptors and the target property. In some cases, a simple linear function can already 
be used to relate one descriptor to a target property, while in other cases more complex, nonlinear, multivariate expressions 
might be necessary. In contemporary QSPR studies, sophisticated methods such as artificial neural networks and support vector 
machines are employed in order to establish a functional dependence between properties and descriptors\cite{chem_rev_2010_110_5714}. 
Once such a dependence has been established, one can easily predict the desired properties for arbitrary molecules (although 
one has to be careful when predicting a property for a molecule which is very different from any molecule of the original 
training set\cite{chempluschem_2012_77_507}). 

\begin{figure}
  \begin{center}
    \includegraphics[scale=0.2]{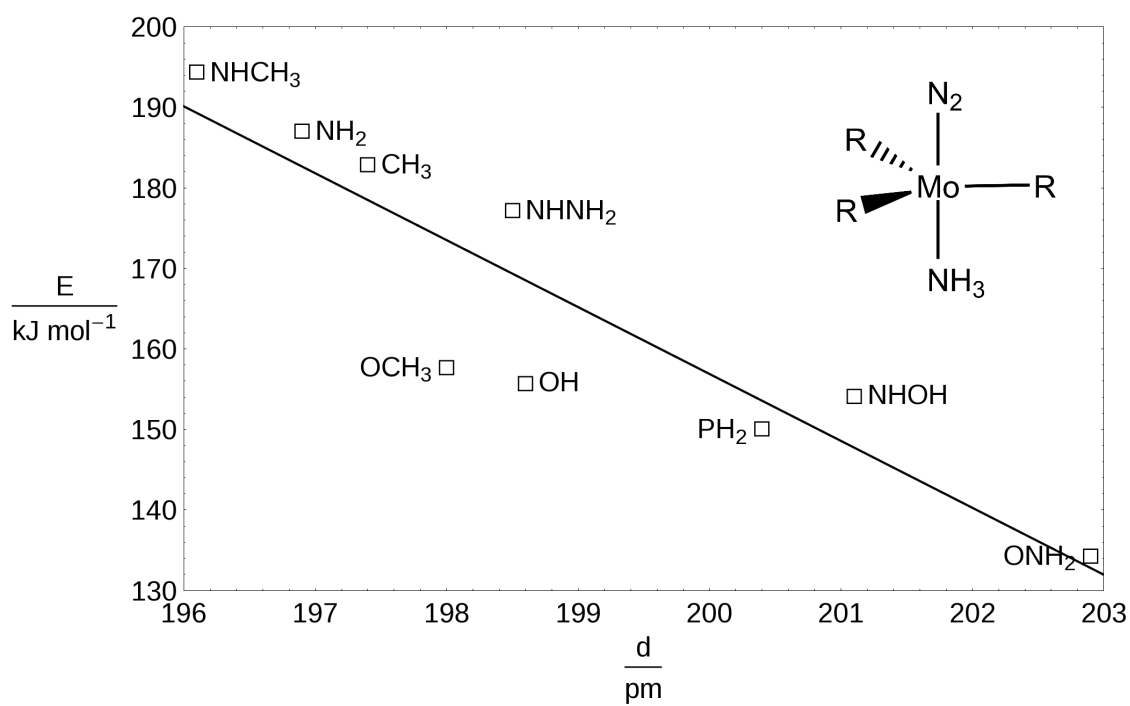}
  \end{center}
  \caption{Dependence of the Mo--N$_2$ binding energy $E$ on the Mo--N bond length $d$ of the trigonal-bipyramidal molybdenum 
complexes with the general structure shown in the top right corner. Calculated data for the molecules are taken from Ref.\ \cite{weym14} and are shown as black squares 
(labeled with the individual residues R). A linear least-squares regression is given by the black line.\label{fig:qsar}}
\end{figure}

Let us illustrate this by a very simple example. Given a set of nine trigonal-bipyramidal molybdenum complexes shown in the 
top right corner of Fig.~\ref{fig:qsar}, we would like to find an expression giving the Mo--N$_2$ binding energy as a function 
of some descriptor(s). As the Mo--N bond length can be expected to depend on the binding energy, it appears natural to choose 
this bond length as a descriptor.

Fig.~\ref{fig:qsar} shows the binding energies as a function of the Mo--N bond length\cite{computdet}. In a first approximation, 
this dependence can be described by a linear function. With a least-squares fit, the binding energy $E$ can be obtained from 
the Mo--N bond length $d$ as
\begin{equation}
 E(d) = 8.31\,{\rm kJ\,mol^{-1}\,pm^{-1}} d + 1818.73\,{\rm kJ\,mol^{-1}}.
\end{equation}
As one can see from Fig.~\ref{fig:qsar}, this relation is overly simplified. Even though the Mo--N bond length can be used 
as a first, rough criterion to estimate the bond strength, it is certainly not the only factor affecting this quantity. This 
can be seen from two complexes in Fig.~\ref{fig:qsar}, which have almost the same bond length (namely, about 198.5\,pm), but 
largely differ in their binding energy. Our calculated expression above would predict the same binding energy for both complexes, 
thus overestimating the value for one complex and underestimating it for the other. A descriptor which could be better suited 
to judge the Mo--N$_2$ binding energy would be the stretching frequency of the Mo--N or N--N bonds. With such an approach, a relationship 
between the N--N stretching frequency and the degree of activation of dinitrogen for a series of Schrock-type nitrogen 
fixating catalysts was established\cite{inorg_chem_2009_48_1638}.

Of course, the development of a robust QSPR for a given quantity is not trivial, and cannot be simply automatized. In particular, 
it is often not \textit{a priori} clear which descriptors should be chosen. Furthermore, a given QSPR is usually only valid 
for a comparatively small subset of chemical compound space, namely for molecules which are very similar to each other in 
terms of their molecular structures. Therefore, these techniques are only of limited applicability for a truly inverse approach, 
to which the entire chemical compound space should be accessible. Moreover, even though it is easy to identify the descriptor 
values necessary for a given desired property once a mathematical relation between this property and its descriptors has been 
found, the construction of a molecule corresponding to a given set of descriptors is not straightforward at all. An approach 
to tackle this problem is denoted as inverse QSAR\cite{j_chem_inf_comput_sci_1998_38_251,j_chem_inf_comput_sci_1998_38_259}. 
In this approach, one aims at constructing a library of chemical compounds similar to a given lead compound (which is known 
to exhibit some favorable property). In this context, chemical similarity is defined as the Euclidean distance of two molecules 
in the space spanned by all descriptors used. Therefore, the more similar the descriptor values are for two compounds, the 
more chemically similar they are expected to be. The generation of new molecules is accomplished by connecting different 
predefined building blocks with each other in order to obtain a chemically reasonable structure. Then, the descriptors of this 
new structure are computed. They can be used in an optimization algorithm such as simulated annealing in order to find structures 
which exhibit a predefined similarity to the lead compound. One can thus identify further compounds worth to be closely 
investigated without the need of screening huge molecular libraries.

\section{Advanced Sampling Techniques}
\label{sec:sampling}

Powerful computer systems and clever optimization algorithms (e.g., simulated annealing\cite{science_1983_220_671}, genetic 
algorithms\cite{parallel_comput_1988_7_65}, and Monte Carlo sampling\cite{j_am_stat_assoc_1949_44_335}) allow us to sample 
a comparatively huge part of chemical space. The application of such algorithms is the most important approach in rational 
compound design. Therefore, we shall also review these approaches, although they are not strictly inverse, since they do not 
directly aim at predicting a structure featuring a predefined set of properties, but rather follow the more traditional 
approach where the properties of a predefined molecular structure are calculated.

\subsection{Rational Design in Solid-State Chemistry}
\label{sec:ssc}

During the early 1990s, Sch\"on and Jansen developed a sophisticated technique for synthesis planning\cite{angew_chem_int_ed_engl_1996_35_1286,
angew_chem_int_ed_2002_41_3746}. They proposed a modular computer program, which searches chemical space by means of a global 
optimization technique and identifies promising structure candidates, that can subsequently be refined and analyzed. After some 
external boundary conditions (e.g., initial number of atoms, external pressure, etc.) have been specified, the algorithm sets up 
an elementary cell. The unit cell vectors (length and orientation) are chosen randomly, as are the positions and composition of 
atoms. Then, all these parameters are globally optimized employing a fast but not very accurate electronic-structure method. The 
final result is then subjected to a more accurate refinement. The global optimization is carried out multiple times, starting with 
different random simulation cells. The resulting collection of promising compounds can then be analyzed in terms of their 
composition and physical properties. Jansen and coworkers have applied their methodology to a range of example systems as, for 
example, noble gas mixtures\cite{ber_bunsenges_phys_chem_1995_99_1148} and binary ionic compounds\cite{comput_mater_sci_1995_4_43}.

\subsection{Inverse Band Structure Approach}
\label{sec:ibsa}

A similar approach has been mentioned in 1994 by Werner \textit{et al.}~who proposed to tailor a band structure reflecting the 
desired properties and to subsequently search for a solid exhibiting this very band structure\cite{phys_rev_lett_1994_72_3851}. 
They termed this problem of finding a solid with a suitable band structure the ``inverse band structure problem''. Unfortunately, 
they could not come up with a working solution to this problem. At this point, we should illustrate how complex this optimization 
problem is. Consider a comparatively simple system, such as the pseudo-binary alloy A$_{0.25}$B$_{0.75}$C with a unit cell of 
128 lattice sites (64 cation sites, occupied by either A or B, and 64 anion sites, all of which are occupied by C). Then, the 
total number of possible configurations is given by the binomial coefficient of 64 and 16, i.e., it is on the order of 
10$^{14}$\cite{nature_1999_402_60}. Needless to say that exhaustive enumeration of such a big search space is simply not possible. 
However, only five years later, Franceschetti and Zunger published a computational procedure which successfully addresses the 
inverse band structure problem\cite{nature_1999_402_60}. Similarly to the approach by Sch\"on and Jansen, they start with an 
elementary cell of a given configuration. They define an atomic configuration in terms of a vector
\begin{equation}
 \sigma = (S_1, S_2, \ldots , S_N),
\end{equation}
where $S_i$ is the atom type at lattice site $i$. A given property $P$ (such as the band gap) is a function of the atomic 
configuration. The goal of the optimization is to find a configuration with one or several predefined target properties 
$P^{\rm (target)}$, i.e., one attempts to minimize the function
\begin{equation}
 O(\sigma) = \sum_{\alpha} \omega_{\alpha} \left| P_{\alpha}(\sigma) - P_{\alpha}^{\rm (target)} \right|,
\end{equation}
where $\omega_{\alpha}$ is the weight assigned to the property $P_{\alpha}$. In the original approach by Franceschetti and Zunger, 
this function is minimized by a simulated annealing algorithm. Atomic configurations are generated by Monte Carlo moves 
(e.g., changing the type of a given atom). Even though this algorithm is able to find the global minimum significantly faster 
than it would be possible with exhaustive enumeration, it still requires sampling of several thousand atomic configurations. 
It is therefore very important to have a fast method at hand to evaluate the properties of a given atomic configuration. 
Franceschetti and Zunger relied on a valence-force-field method to optimize the positions of the atoms of a given atomic 
configuration\cite{phys_rev_b_1995_51_10795}, and utilized a semiempirical Hamiltonian, where each atom is replaced by a 
pseudopotential\cite{phys_rev_b_1995_51_17398}, in order to evaluate the properties. Furthermore, they applied a special 
technique for the diagonalization of the Hamiltonian, which focuses on the few eigenvalues around the band gap\cite{j_chem_phys_1994_100_2394}.

Franceschetti and Zunger predicted\cite{nature_1999_402_60} the configuration of an Al$_{0.25}$Ga$_{0.75}$As alloy with a 
maximum band gap. Recently, Zunger and coworkers studied a quaternary (In,Ga)(As,Sb) semiconductor\cite{phys_rev_lett_2008_100_186403}. 
The original approach has continuously been further developed and improved. For example, it is also possible to employ a 
genetic algorithm instead of simulated annealing\cite{j_comput_phys_2005_208_735,phys_rev_lett_2006_97_046401}, which generally 
leads to a faster convergence.

We should note here that the {\it general} method of finding the optimal species with regard to a given property in a subset 
of chemical space is also employed and further developed by many other groups, although they do not specifically aim at an 
inverse approach. For example, N\o{}rskov and coworkers have identified the 20 most stable four-component alloys of 32 
transition, noble and simple metals in both face-centered cubic and body-centered cubic structures\cite{phys_rev_lett_2002_88_255506}. 
For this, they searched a subspace of no less than 192'016 compounds by means of an evolutionary algorithm\cite{phys_rev_lett_2002_88_255506}. 
Given the limited space here, we refer the reader to Refs.\ \cite{nature_chem_2009_1_37,j_mater_sci_2012_47_7317,mrs_bulletin_2006_31_681,mrs_bulletin_2010_35_693,phys_rev_lett_2010_105_196403,
appl_phys_lett_mater_2013_1_011022,inorg_chem_2011_50_656} for further examples of computational materials design and to Refs.\
\cite{stru13,truh13} for an example of environment-enhanced reactivity provided by inverse solvent design
of a condensed phase reaction.

The approach by Franceschetti and Zunger is specifically tailored for the design of solid-state materials, and its extension 
to the design of isolated molecules is not necessarily straightforward. In the above procedure, the total number of atoms is 
defined from the outset, as a certain unit cell with a given number of lattice sites is defined. However, for an isolated 
molecule, this approach is not valid. One would therefore have to abandon this constraint and introduce different ones. For 
example, by means of simple valence rules, one could define the total number of atoms coordinating to a given atom (i.e., 
the number of nearest neighbors). Furthermore, in the design of solid-state materials one can apply periodic boundary conditions, 
and therefore in principle study a crystal of infinite size. For an isolated molecule, however, this is not possible, and 
one therefore encounters the problem of unsaturated valencies.

\subsection{Linear Combination of Atom-Centered Potentials}
\label{sec:lcacp}

One of the more daunting problems in the field of rational design is the fact that individual molecules represent discrete 
points in chemical space, whereas most optimization methods are developed to work with continuous functions. In 2006, Beratan, 
Yang and coworkers addressed this problem\cite{j_am_chem_soc_2006_128_3228}. They started from the well-known fact that 
within the framework of density functional theory any molecular system is characterized by the external potential 
$\upsilon(\boldsymbol{r})$\cite{phys_rev_1964_136_864}, which represents the attraction between electrons and nuclei, 
and the total number of electrons. They then treated this external potential as a continuous variable to be optimized. 
Yang \textit{et al.}~even presented a functional of external potentials together with a corresponding variational 
principle\cite{phys_rev_lett_2004_92_136404}. One major issue with this approach is that not all external potentials 
correspond to a molecule, even though all molecules map unequivocally to a given $\upsilon(\boldsymbol{r})$. An 
external potential which corresponds to a chemical structure is called C-representable. Beratan, Yang and coworkers very 
elegantly solved this problem of C-representability by expanding the potential $\upsilon(\boldsymbol{r})$ into a set of 
atomic potentials\cite{j_am_chem_soc_2006_128_3228},
\begin{equation}
 \upsilon(\boldsymbol{r}) = \sum_{A, \boldsymbol{R}}b_A^{(\boldsymbol{R})}\upsilon_A^{(\boldsymbol{R})}(\boldsymbol{r}),
\end{equation}
where $b_A^{(\boldsymbol{R})}$ is the expansion coefficient of the potential term $\upsilon_A^{(\boldsymbol{R})}(\boldsymbol{r})$, 
which denotes the potential of atom $A$ at position $\boldsymbol{R}$. This approach has been called ``linear combination of 
atomic potentials'' (LCAP), and it constrains the external potential to be necessarily C-representable. The optimization problem 
then reduces to finding the best set of expansion coefficients for these atomic potentials. It is important to note that not 
only individual atoms can be represented by such atomic potentials but a collection of such potentials can be contracted to 
represent e.g., a functional group,
\begin{equation}
 \upsilon_A^{(\boldsymbol{R})}(\boldsymbol{r}) = \sum_B b_B^{(\boldsymbol{R})}\upsilon_B^{(\boldsymbol{R})}(\boldsymbol{r}),
\end{equation}
and can be optimized as a whole. In order to correspond to an actual molecule, the coefficients $b_A^{(\boldsymbol{R})}$ must 
be either 0 or 1 denoting either the absence or the presence of the atom or functional group, respectively, represented by 
$\upsilon_A^{(\boldsymbol{R})}(\boldsymbol{r})$. However, since the expansion coefficients are treated as continuous variables 
during the optimization, one can find non-integer values for them. In this case, rounding to the nearest integer is necessary. 
While this approach of continuous optimization is rather new in quantum chemistry, Zunger and coworker pointed out that Bends\o{}e 
and Kikuchi had already presented a similar approach in the context of materials design\cite{comp_methods_appl_mech_eng_1988_71_197}.

In their original publication\cite{j_am_chem_soc_2006_128_3228}, Beratan, Yang and coworkers investigated as a proof of 
concept a rather simple problem, constructed from two sites (fixed in space), for which they searched for the best potential 
such that the resulting hyperpolarizability is maximized. In doing so, they started from a small library of six different 
chemical substituents modeled by different potentials, and investigated how the optimal potential is composed of these ``atomic'' 
potentials. They finally found that H$_2$S$_2$ has the largest hyperpolarizability of all molecules present in the small 
chemical space considered (six substituents at two sites give a total of 2\textsuperscript{6} = 64 possible molecules) which 
was in agreement with the results found by exhaustive enumeration. With the same methodology, Keinan {\it et al.}~successfully 
optimized the first hyperpolarizability of porphyrin-based nonlinear optical materials\cite{j_phys_chem_a_2008_112_12203}. 
d'Avezac and Zunger recently published results showing that the approach by Yang and coworkers is very efficient for optimizing 
the atomic configurations of alloys\cite{j_phys_condens_matter_2007_19_402201}.

In recent years, the group of Yang continuously developed and improved the method further. In 2007, they implemented it into 
an AM1 semiempirical framework\cite{j_phys_chem_a_2007_111_176}. In 2008, a gradient-directed Monte Carlo approach, which 
is able to ``jump'' between chemical structures, and thus to overcome barriers between local minima, was presented\cite{j_chem_phys_2008_129_064102}. 
The approach was also implemented within a tight-binding framework\cite{j_chem_phys_2008_129_044106}. Furthermore, these 
authors studied different optimization procedures and found that a combination of their LCAP approach with a best-first search 
algorithm (BFS, an algorithm searching a graph by stepwise expanding the most promising node\cite{pear84}) improves the 
performance of the optimization\cite{j_chem_phys_2008_129_174105}.

The LCAP and related approaches have reached a state of maturity which allows them to be applied to rather complex problems\cite{j_phys_chem_a_2008_112_12203}. 
However, the position of the potentials is fixed in space (and only optimized in the course of a standard structure optimization), 
which restricts the space of chemical compounds studied. This bears the danger of overlooking a good candidate, simply because 
it is not in the sampled space. This problem could be remedied by employing an extremely large library and allowing potentials 
at many different sites. However, this would also dramatically increase the computational cost of the resulting optimization 
problem. Nevertheless, as computer systems become more and more powerful, the LCAP will certainly be applied to increasingly 
complex problems.

\subsection{Alchemical Potentials}
\label{sec:alchpot}

Another interesting approach to convert chemical space into a continuous space has been developed by von Lilienfeld {\it et al.} 
Starting again from the Hohenberg--Kohn theorem\cite{phys_rev_1964_136_864}, von Lilienfeld {\it et al.}~noted that any molecular 
property $O$ can be written as a functional of the nuclear charge distribution $Z(\boldsymbol r)$ and as a function of the total 
electron number $N_{\rm el}$\cite{phys_rev_lett_2005_95_153002}, i.e., $O \equiv O[Z(\boldsymbol r)](N_{\rm el})$. Pictorially 
speaking, molecules are superpositions of electron and nuclear charge distributions. The electron density, however, is (up to 
a constant factor) fully determined by the total electron number and the external potential, which itself is a functional of the 
nuclear charge distribution. Rational compound design can thus be described as an optimization problem within the space spanned 
by nuclei and electrons, which leads us to the minimization problem\cite{phys_rev_lett_2005_95_153002}
\begin{equation}
 \min_{N_{\rm el}, Z(\boldsymbol r)} \left| O[Z(\boldsymbol r)](N_{\rm el}) - O_0 \right|^2,
\end{equation}
where $O_0$ is the desired target property. A continuous optimization in the space of electrons and nuclei implies fractional 
electronic and nuclear charges. At this point, we should note that some controversy concerning fractional electrons emerged some 
years ago and is still not fully resolved (see, e.g., Ref.~\cite{science_2008_321_792}). In practice, in von Lilienfeld's 
approach fractional atomic and electronic charges are generated by means of a suitably parametrized effective core potential 
(ECP)\cite{int_j_quantum_chem_2013_doi_101002qua24375}. Therefore, there are no explicit fractional electrons occurring in the 
computation, but they are rather implicitly dealt with in the ECP.

In order to efficiently search the space spanned by electrons and nuclei, von Lilienfeld {\it et al.}~worked out a mathematical 
expression for the derivative of the total electronic energies with respect to the nuclear charge distribution\cite{phys_rev_lett_2005_95_153002}. 
This has subsequently become known as the nuclear chemical potential $\mu_n(\boldsymbol{r})$ (in analogy to the electronic chemical 
potential), which, for a discrete nuclear charge distribution and perturbation theory to the first order, is found to be the 
electrostatic field $E^{(1)}(\boldsymbol{r})$,
\begin{equation}
 \mu_n(\boldsymbol{r}) \approx E^{(1)}(\boldsymbol{r}) = -\int \frac{\rho(\boldsymbol{r}')}{\left| \boldsymbol{r} - \boldsymbol{r}' \right|}{\rm d}\boldsymbol{r}' + \sum_I\frac{Z_I}{\left| \boldsymbol{r} - \boldsymbol{r}_I \right|},
\end{equation}
where the integration is over the entire space. The nuclear chemical potential is a function of space and is related to the 
proton affinity\cite{phys_rev_lett_2005_95_153002}. At the position of a nucleus, it measures the tendency of this nucleus to 
``mutate'' into a different type\cite{phys_rev_lett_2005_95_153002}. At these specific points, the nuclear chemical potential 
is usually called the alchemical potential\cite{phys_rev_lett_2005_95_153002}.

Von Lilienfeld and Tuckerman demanded a rigorous description of chemical space\cite{j_chem_phys_2006_125_154104}. They found 
that such a description is tightly connected to grand-canonical ensemble theory and, hence, could benefit from earlier developments. 
It was soon recognized\cite{j_chem_phys_2006_125_154104} that molecular properties such as the potential energy represent a 
state function within the space of $Z(\boldsymbol r)$ and $N_{\rm el}$, and hence, any two points in this space can be connected 
by an order parameter, which is usually denoted as $\lambda$. This parameter can be used to interpolate between any two molecules 
(similar to the thermodynamic integration) and to study the variation of a given property as one molecule is continuously changed 
into the other\cite{j_chem_phys_2006_125_154104,j_chem_theory_comput_2007_3_1083,j_chem_phys_2009_131_164102,j_chem_phys_2010_133_084104}. 
In such studies, von Lilienfeld found that the path connecting any two molecules is generally not linear\cite{j_chem_phys_2009_131_164102}. 
This makes the prediction of properties of unknown molecules based on results of known compounds very difficult. A general 
method of linearizing these connection paths has yet to be found, and von Lilienfeld even offers a prize (the equivalent of an 
ounce of gold) to the person finding a solution to this complicated problem\cite{int_j_quantum_chem_2013_doi_101002qua24375}.

So far, this approach has not been applied to many problems yet. As a proof of principle von Lilienfeld {\it et al.}\cite{phys_rev_lett_2005_95_153002} 
designed a nonpeptidic anticancer drug candidate by optimizing the atoms constituting the peptide unit in a peptidic starting 
structure (even though this peptidic structure already showed strong antitumoral activity, it would be cleaved {\it in vivo} 
by proteases, such that it can essentially not be employed as drug). In 2007, Marcon {\it et al.}~showed that the energy of 
the highest occupied molecular orbital in benzene and its BN-doped derivatives can be precisely tuned\cite{j_chem_phys_2007_127_064305}.

\section{Mode- and Intensity-Tracking}
\label{sec:tracking}

In this section we consider subspace iteration techniques that we have devised to target molecular vibrations, for which one
can either provide a simple structural distortion as a first guess (Mode-Tracking) or which pick up intensity in some spectroscopic set-up (Intensity-Tracking).
Both Mode- and Intensity-Tracking are rather different from all of the above-mentioned approaches in the sense that they do not 
aim for the rational design of a molecular compound or material. However, they belong to the realm of inverse methods as they 
turn the usual procedure in quantum chemical calculations upside-down\cite{new_j_chem_2007_31_818}. More specifically, 
Mode-Tracking\cite{j_chem_phys_2003_118_1634} allows for the iterative, but direct optimization of qualitatively predefined 
(local or collective) molecular vibrations without the need of first calculating the entire Hessian matrix (as would be the 
case in the standard approach). Similarly, Intensity-Tracking enables one to iteratively converge a spectrum based on the 
most intense vibrations starting from intensity carrying structural distortions\cite{chimia_2009_63_270,j_chem_phys_2008_129_204103,j_chem_phys__2009_130_064105,chemphyschem_2009_10_2049}. 
Thus, these two methods directly target selected molecular vibrations of some predefined characteristic and allow one to obtain these at lower computational cost than 
with the traditional methodologies.

The standard approach to calculate the $k$-th vibrational frequency of a molecule is to solve the harmonic eigenvalue problem
\begin{equation}\label{eq:hq}
 \boldsymbol{H}^{\rm (m)} \boldsymbol{q}_k = \lambda_k \boldsymbol{q}_k,
\end{equation}
where the eigenvector $\boldsymbol{q}_k$ is called the $k$-th normal mode and the eigenvalue $\lambda_k$ is proportional to 
the square of the $k$-th vibrational angular frequency. $\boldsymbol{H}^{\rm (m)}$ is the mass-weighted Hessian matrix (in 
Cartesian coordinates), the entries of which are given by\cite{snf}
\begin{equation}
 H^{\rm (m)}_{I\alpha,J\beta} = \frac{1}{\sqrt{M_I M_J}}\left( \frac{\partial^2 E_{{\rm el}}}{\partial r_{I\alpha} \partial r_{J\beta}} \right),
\end{equation}
where $M_I$ and $\boldsymbol{r}_I$ denote the mass and spatial position of nucleus $I$ ($\alpha \in \{x, y, z\}$), respectively. 
In the following, we will abandon these two-component subscripts in favor of only one subscript ranging from 1 to 3$N$ 
written in lower case. The idea of Mode-Tracking is to selectively calculate certain normal modes without first computing the 
entire Hessian matrix, which is the most time-consuming step of a standard vibrational analysis\cite{j_chem_phys_2003_118_1634}. 
It is usually straightforward to create a molecular distortion as an approximation for the desired vibration. One can therefore 
start with a guess mode
\begin{equation}
 \boldsymbol{b} = \sum_{i=1}^{3N}b_i\boldsymbol{e}_i^{\rm (m)},
\end{equation}
where the sum runs over all Cartesian coordinates of the $N$ nuclei, $b_i$ is the displacement magnitude of the mode in the 
$i$-th component, and $\boldsymbol{e}_i^{\rm (m)}$ are the mass-weighted nuclear Cartesian basis vectors. Taking $\boldsymbol{b}$ 
as a first approximation to the desired normal mode $\boldsymbol{q}_k$, one can calculate the $k$-th element of the left-hand 
side of Eq.~(\ref{eq:hq}) as
\begin{equation}
 \sigma_k = \left( \boldsymbol{H}^{\rm (m)} \boldsymbol{q} \right)_k = \sum_j \frac{1}{\sqrt{M_j M_k}} \frac{\partial^2 E_{\rm el}}{\partial r_j \partial r_k}b_j = \frac{1}{\sqrt{M_k}} \frac{\partial^2 E_{\rm el}}{\partial r_k \partial \boldsymbol{b}}.
\end{equation}
The last equality means that $\sigma_k$ can be calculated by computing the directional derivative of the nuclear gradient along 
the collective distortion $\boldsymbol{b}$, which can  be conveniently done in a semi-numerical fashion\cite{j_chem_phys_2003_118_1634}. 
In Mode-Tracking, the target mode is iteratively improved by a Davidson-type subspace iteration method\cite{j_chem_phys_2003_118_1634}. 
In every iteration $l$, a new vector $\boldsymbol{b}^{(l)}$ is generated, and concomitantly, also a new vector $\boldsymbol{\sigma}^{(l)}$ 
is obtained. The $i$ vectors generated up to iteration $i$ can be assembled into the matrices $\boldsymbol{B}^{(i)}$ and 
$\boldsymbol{\Sigma}^{(i)}$, respectively, to produce the matrix
\begin{equation}
 \boldsymbol{\tilde H}^{{\rm (m}, i)} = \boldsymbol{B}^{(i)T}\boldsymbol{H}^{\rm (m)} \boldsymbol{B}^{(i)} = \boldsymbol{B}^{(i)T}\boldsymbol{\Sigma}^{(i)},
\end{equation}
i.e., it is not necessary to calculate the full Hessian matrix $\boldsymbol{H}^{\rm (m)}$. We then solve the small eigenvalue 
problem
\begin{equation}
 \boldsymbol{\tilde H}^{{\rm (m}, i)}\boldsymbol{u}^{(i)} = \boldsymbol{\rho}^{(i)}\boldsymbol{u}^{(i)},
\end{equation}
where the matrices $\boldsymbol{\rho}^{(i)}$ and $\boldsymbol{u}^{(i)}$ contain the the $i$-th approximations to the target 
eigenvalue $\lambda_k$ and target normal mode $\boldsymbol{q}_k$, respectively. The best approximations, which we will denote 
as $\rho_s^{(i)}$ and $\boldsymbol{u}^{(i)}_s$, are selected from $\boldsymbol{\rho}^{(i)}$ and $\boldsymbol{u}^{(i)}$, and 
the residuum vector
\begin{equation}
 \boldsymbol{r}^{(i)}_s = \sum_{l=1}^iu^{(i)}_{s,l}\left( \boldsymbol{\sigma}^{(l)} - \rho_s^{(i)}\boldsymbol{b}^{(l)} \right)
\end{equation}
is calculated. From this residuum vector, a new basis vector $\boldsymbol{b}^{(i+1)}$ is generated,
\begin{equation}
 \boldsymbol{b}^{(i+1)} = \boldsymbol{X}^{(i)}\boldsymbol{r}^{(i)}_s,
\end{equation}
where $\boldsymbol{X}^{(i)}$ is a preconditioner. The convergence characteristics of the Mode-Tracking algorithm strongly 
depend not only on the initial guess mode, but also on the particular choice of $\boldsymbol{X}^{(i)}$. However, experience 
shows that also very simple choices for the preconditioner (e.g., a unit matrix) facilitate fast convergence\cite{phys_chem_chem_phys_2004_6_4621}.

The Mode-Tracking approach has successfully been applied in studies of a large [(Ph$_3$PAu)$_6$C]$^{2+}$ cluster\cite{j_comput_chem_2004_25_587}, 
molecular wires (i.e., long carbon chains)\cite{j_phys_chem_a_2004_108_2053}, adsorbate vibrations\cite{surf_science_2006_600_1891}, 
and tetrameric methyl lactate clusters\cite{angew_chem_2006_118_3518}. Furthermore, in 2008, Herrmann {\it et al.}~integrated 
the Mode-Tracking methodology into a QM/MM framework\cite{j_comput_chem_2008_29_2460}. Recently, Kovyrshin and Neugebauer\cite{j_chem_phys_2010_133_174114,chem_phys_2011_391_147} presented a similar tracking methodology for the selective 
calculation of electronic excitations in the framework of time-dependent DFT.

Intensity-Tracking generally works according to the same subspace-iteration principle explained above for Mode-Tracking. However, the approach does 
not aim at refining a given user-defined vibration, but to yield those normal modes which are associated with highest intensities 
(for a given vibrational spectroscopy). Therefore, the most important difference to Mode-Tracking consists in the choice of 
the starting guess mode. A suitable guess mode is a molecular distortion that attains intensity. Such intensity carrying modes 
can generally be derived from an eigenvalue problem defined for the particular spectroscopy under consideration. 
This guess depends strongly on the kind of vibrational spectroscopy for which the highest-intensity 
modes are to be obtained. 
We have developed Intensity-Tracking for conventional infrared spectroscopy\cite{j_chem_phys__2009_130_064105}, Raman and 
Raman optical activity spectroscopy\cite{chemphyschem_2009_10_2049}, as well as resonance Raman spectroscopy\cite{j_chem_phys_2008_129_204103}.

Mode- and Intensity-Tracking demonstrate that subspace iteration methods are a feasible way to iteratively solve an inverse problem\,---\,here, 
to optimize normal modes from pre-defined molecular distortions.

\section{Gradient-driven Molecule Construction\label{GdMC-sec}}

The ultimate criterion for the stability of a given molecule is the Gibbs (or Helmholtz) free energy. However, in (chemical) processes involving
energy changes much larger than the thermal energy RT  the entropic (and temperature) contributions 
may be neglected as the change in free energy is then dominated by the change in electronic energy. This situation simplifies the design problem as it
is governed by the electronic structure of the reactants in the chemical process under study. 
The electronic energy difference is calculated from the electronic energies of the reactants. The latter
are determined as stationary points on the Born--Oppenheimer potential energy surface $E_{el}$ of Eq.\ (\ref{elS-Gl}), which is the mathematical object that relates all electronic
energies calculated for fixed nuclear frameworks of a given number of atomic nuclei (and electrons). Consequently, a characteristic of stable reactants
{\it and} of first-order transition structures is a vanishing first derivative of the electronic energy with respect to all nuclear coordinates, i.e.,
a vanishing geometry gradient. Therefore, the geometry gradient is the central property for determining the stability of a molecule. 
This fact is utilized in {\it Gradient-driven Molecule Construction} (GdMC) \cite{weym14}, a new concept that
we have proposed for the rational design of functional molecules in 2012 \cite{mrs_proceedings_2013_1524_doi}.

Given some desired structural feature within a molecule or a molecular assembly, the length of the gradient on the atoms in the isolated structural feature, i.e., the 
length of the fragment's gradient,
\begin{equation}\label{eq:abgrad}
 |\nabla_{\rm frag} E_{\rm el}| = \sqrt{ \sum_{I\in {\rm frag}} \left[ \left(\frac{\partial E_{\rm el}}{\partial r_{I, x}} \right)^2 + \left( \frac{\partial E_{\rm el}}{\partial r_{I, y}} \right)^2 + \left( \frac{\partial E_{\rm el}}{\partial r_{I, z}} \right)^2\right]},
\end{equation}
with $r_{I, \alpha}$ ($\alpha\in\{x,y,z\}$) being a nuclear Cartesian coordinate,
is in general different from zero and not even small. The gradient components of this fragment can become vanishingly small upon embedding of the fragment in
an appropriate molecular environment. Thus, the vanishing gradient criterion of GdMC can be exploited for the search of molecular scaffolds that fulfill a certain chemical 
purpose provided that this purpose can be pre-defined in terms of a molecular structure.
For instance, one may ask the question which chelate ligand can stabilize a molecular fragment, such as a transition metal binding a small inert molecule,
by designing the scaffold of the chelate ligand in such a way that the geometry gradient at each nucleus in the compound system vanishes. 

The structure of the fragment may even be chosen in such a way that some bonds are elongated compared to the isolated reactants, from which the fragment is formed, so that the 
situation in the fragment represents some degree of activation with respect to the isolated reactants.
For example, when a dinitrogen molecule binds to a transition metal center, its triple bond might remain more or less intact upon coordination and thus
the N$\equiv$N bond length hardly changes. On the other hand, upon binding of the dinitrogen ligand electron transfer can lead to an activation in such a way
that a diazenoid double bond or even a hydrazinoid single bond is formed with the respective characteristic bond lengths \cite{sellmann1,sellmann1}.
Hence, prototypical bond lengths and angles can be pre-defined as necessitated by the overall chemical process
of which the species to be designed is an intermediate. Hence, a design may be carried out for each node of a reaction network and mutual
stability dependencies may also need to be fulfilled resulting in a highly complex optimization problem. However, different optimization
approaches are possible within GdMC, and we have explored some of them in our study on dinitrogen binding molybdenum complexes \cite{weym14}.
In a first approach, the Schrock complex, which is known to bind molecular nitrogen (see Fig.~\ref{fig:gdmc}), was taken 
as an example of an existing nitrogen-binding complex that could be used as a known reference to validate the approach. In one optimization procedure,
a range of model complexes was deduced from this compound in which a 
central Mo--N$_2$ fragment was taken from the full Schrock complex 
but where the ligand sphere was modeled by the variable substituents R$_1$ and R$_2$. Then, the atoms or 
functional groups for R$_1$ and R$_2$ were searched such that the geometry gradient on all atoms of the complex was minimal, 
leading to a stable complex featuring the predefined bond lengths mentioned above.

\begin{figure}
  \label{fig:gdmc}
  \begin{center}
    \includegraphics[scale=0.1]{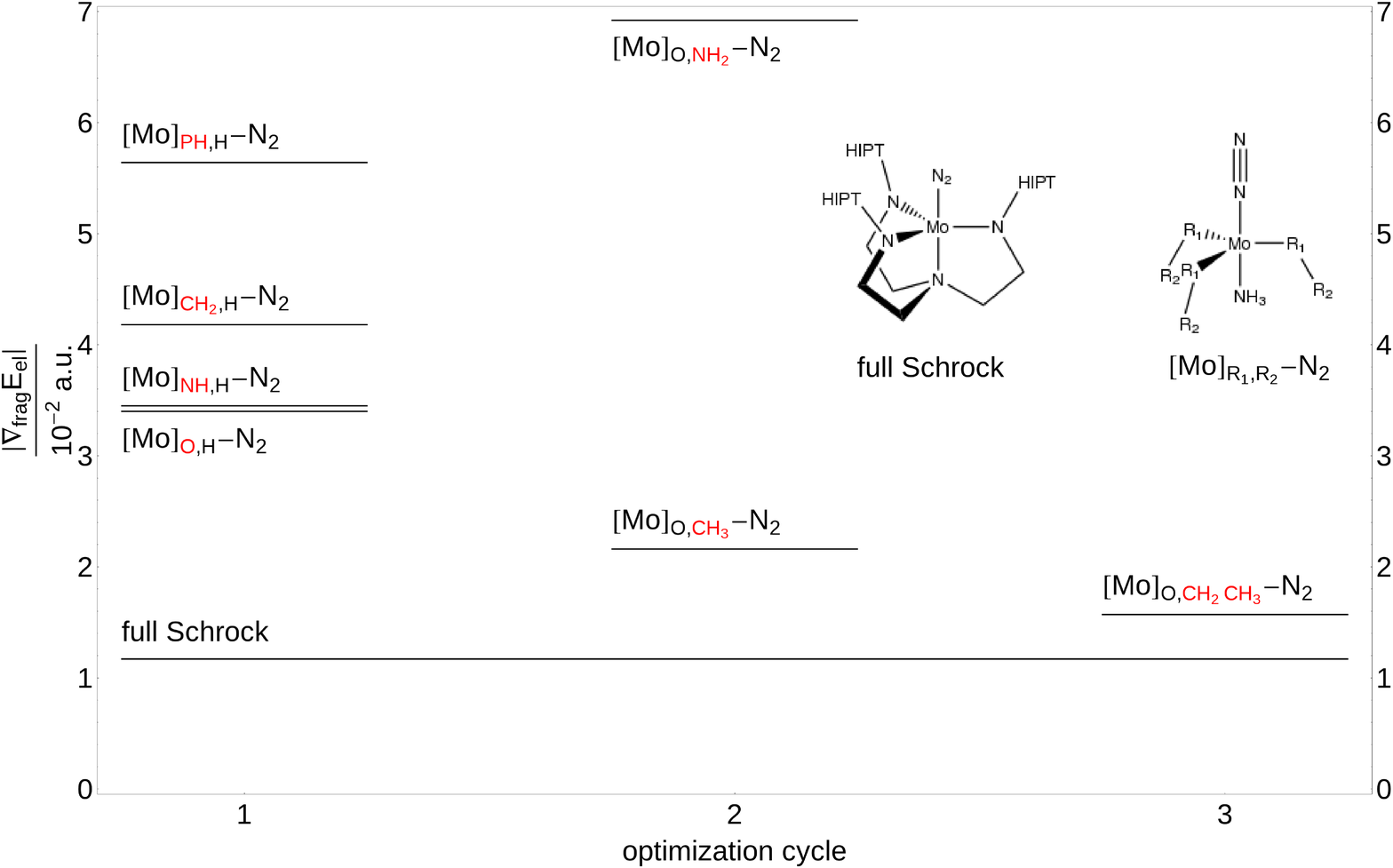}
  \end{center}
  \caption{Stepwise construction of a nitrogen-binding transition metal complex following the concept of gradient-driven molecule construction. The data for this figure have been taken from Ref.\ \cite{weym14} (see this reference for technical details).
\label{fig:gdmc}}
\end{figure}

This procedure is illustrated in Fig.~\ref{fig:gdmc}. In a first optimization cycle, different atoms (oxygen, nitrogen, carbon, 
and phosphorus) were placed at R$_1$ and all valencies were saturated with hydrogen atoms. An analysis of the resulting geometry 
gradient revealed that oxygen is best suited for minimizing the geometry gradient. One can further reduce the gradient by varying 
R$_2$ in further optimization cycles. We found that with an ethyl group for R$_2$, our model complex featured a gradient which 
was already very close to the reference gradient of the original Schrock complex.

So far we have taken only the first steps within the GdMC approach, but we expect that it can be turned into a black-box inverse as well as high-throughput 
screening tool for the design of molecules with well-defined structural features in one of its fragment structures.

\section{Conclusion and Outlook}
\label{sec:conclusion}

In this work we have aimed at providing an overview on inverse approaches in quantum chemistry for the direct rational design of molecular properties and function.
The rational design of molecules and materials remains a challenging task, despite the many achievements accomplished in this field 
over the past decades. The most advanced contemporary methods all rely on sophisticated optimization algorithms and powerful 
computer hardware. Therefore, future advances in these areas will enable us to tackle increasingly complicated 
problems. Apart from new theoretical and algorithmic developments in inverse quantum chemistry, the possibility to carry out quantum chemical 
calculations (almost) instantaneously, i.e., in real time and thus interactively \cite{haag13} will further 
enhance our capabilities (with respect to automated procedures as well as with respect to manually steered quantum-chemical
design protocols \cite{marti2009,haag2011,bosson2011,bosson2012,haag2014}). 
Clearly, massive screening approaches are now feasible and help us better understand the relation between
molecular structure and property. This knowledge should put us in a position to devise improved or novel inverse methods, which can be set up to serve
certain well-defined design goals.

\section*{Acknowledgments}

This work has been supported by the Swiss National Science Foundation SNF (project 200020\_144458/1).

\providecommand{\refin}[1]{\\ \textbf{Referenced in:} #1}


\begin{thebibliography}{100}

\bibitem{dyks05}
\emph{{Theory and Applications of Computational Chemistry: The First Forty
  Years}};
\newblock Dykstra,~C.~E.; Frenking,~G.; Kim,~K.~S.; Scuseria,~G.~E., Eds.;
\newblock Elsevier Science Publishers: Amsterdam, \textbf{2005}.

\bibitem{helg00}
Helgaker,~T.; J{\o}rgensen,~P.; Olsen,~J. \emph{{Molecular Electronic-Structure
  Theory}};
\newblock John Wiley \& Sons: Chichester, \textbf{2000}.

\bibitem{parr89}
Parr,~R.~G.; Yang,~W. \emph{{Density-Functional Theory of Atoms and
  Molecules}};
\newblock Oxford University Press: New York, \textbf{1989}.

\bibitem{cohe09}
Cohen-Tannoudji,~C.; Diu,~B.; Lalo\"{e},~F. \emph{{Quantum Mechanics}};
\newblock John Wiley \& Sons: New York, \textbf{1977}.

\bibitem{geophys_j_int_2007_171_1118}
Campman,~X.; Riyanti,~C.~D. \emph{Geophys. J. Int.} \textbf{2007}, \emph{171},
  1118--1125.

\bibitem{engl97}
\emph{{Inverse Problems in Medical Imaging and Nondestructive Testing:
  Proceedings of the Conference in Oberwolfach, Federal Republic of Germany,
  February 4\,--\,10, 1996}};
\newblock Engl,~H.~W.; Louis,~A.~K.; Rundell,~W., Eds.;
\newblock Springer: Wien, \textbf{1997}.

\bibitem{computer-aided_design_2005_37_509}
Iyer,~N.; Jayanti,~S.; Lou,~K.; Kalyanaraman,~Y.; Ramani,~K.
  \emph{Computer-Aided Design} \textbf{2005}, \emph{37}, 509--530.

\bibitem{chad77}
Chadan,~K.; Sabatier,~P.~C. \emph{{Inverse Problems in Quantum Scattering
  Theory}}, 2nd ed.;
\newblock Springer: New York, \textbf{1989}.

\bibitem{int_j_quantum_chem_2009_109_2456}
Karwowski,~J. \emph{Int. J. Quantum Chem.} \textbf{2009}, \emph{109},
  2456--2463.

\bibitem{j_chem_inf_comput_sci_2003_43_374}
Ertl,~P. \emph{J. Chem. Inf. Comput. Sci.} \textbf{2003}, \emph{43}, 374--380.

\bibitem{med_res_rev_1996_16_3}
Bohacek,~R.~S.; McMartin,~C.; Guida,~W.~C. \emph{Med. Res. Rev.} \textbf{1996},
  \emph{16}, 3--50.

\bibitem{nature_2004_432_824}
Dobson,~C.~M. \emph{Nature} \textbf{2004}, \emph{432}, 824--828.

\bibitem{chem_soc_rev_1973_2_21}
Hall,~G.~G. \emph{Chem. Soc. Rev.} \textbf{1973}, \emph{2}, 21.

\bibitem{trub87}
P\"{o}schel,~J.; Trubowitz,~E. \emph{{Inverse Spectral Theory}};
\newblock Academic Press, Inc.: Orlando, \textbf{1987}.

\bibitem{name}
 Note that one often encounters the English transliterations
  ``Ambartsumian'' and ``Hambardzumyan'' for the last name. In this paper,
  however, we adopt the transliteration as reported in Ambarzumian's
  paper\cite{zeitschrift_f_phys_1929_53_690}.

\bibitem{zeitschrift_f_phys_1929_53_690}
Ambarzumian,~V. \emph{Zeitschrift f. Phys.} \textbf{1929}, \emph{53}, 690--695.

\bibitem{proc_cam_phil_soc_1926_23_44}
Ince,~E.~L. \emph{Proc. Cam. Phil. Soc.} \textbf{1926}, \emph{23}, 44--46.

\bibitem{rayl45}
Rayleigh,~J. W.~S. \emph{{The Theory of Sound}}, 2nd ed.;
\newblock Dover: New York, \textbf{1945}.

\bibitem{acta_mathematica_1946_78_1}
Borg,~G. \emph{Acta Mathematica} \textbf{1946}, \emph{78}, 1--96.

\bibitem{nat_mater_2004_3_838}
Jansen,~M.; Sch\"on,~J.~C. \emph{Nat. Mater.} \textbf{2004}, \emph{3}, 838.

\bibitem{phys_rev_lett_2005_95_153002}
von Lilienfeld,~O.~A.; Lins,~R.~D.; Rothlisberger,~U. \emph{Phys. Rev. Lett.}
  \textbf{2005}, \emph{95}, 153002.

\bibitem{j_mol_spectrosc_1973_46_25}
Albritton,~D.~L.; Harrop,~W.~J.; Schmeltekopf,~A.~L.; Zare,~R.~N. \emph{J. Mol.
  Spectrosc.} \textbf{1973}, \emph{46}, 25--36.

\bibitem{j_mol_spectrosc_1974_51_321}
Kirschner,~S.~M.; Watson,~J. K.~G. \emph{J. Mol. Spectrosc.} \textbf{1974},
  \emph{51}, 321--333.

\bibitem{j_chem_phys_1995_102_2282}
Ho,~T.; Rabitz,~H.; Choi,~S.~E.; Lester,~M.~I. \emph{J. Chem. Phys}
  \textbf{1995}, \emph{102}, 2282--2285.

\bibitem{j_chem_phys_1995_103_9713}
Zhang,~D.~H.; Light,~J.~C. \emph{J. Chem. Phys.} \textbf{1995}, \emph{103},
  9713--9720.

\bibitem{comp_phys_comm_2000_128_622}
Pashov,~A.; Jastrz\k{e}bski,~W.; Kowalczyk,~P. \emph{Comp. Phys. Comm.}
  \textbf{2000}, \emph{128}, 622--634.

\bibitem{j_mol_spectrosc_1975_56_93}
Kosman,~W.~M.; Hinze,~J. \emph{J. Mol. Spectrosc.} \textbf{1975}, \emph{56},
  93--103.

\bibitem{j_chem_phys_1986_85_5151}
Hamilton,~I.~P.; Light,~J.~C.; Whaley,~K.~B. \emph{J. Chem. Phys.}
  \textbf{1986}, \emph{85}, 5151--5157.

\bibitem{chem_phys_lett_1996_252_195}
Wu,~Q.; Zhang,~J. Z.~H. \emph{Chem. Phys. Lett.} \textbf{1996}, \emph{252},
  195--200.

\bibitem{poli02}
Politzer,~P.; Murray,~J.~S. \emph{Theor. Chem. Acc.} \textbf{2002}, \emph{108},
  134--142.

\bibitem{poli11}
Murray,~J.~S.; Politzer,~P. \emph{WIREs Comput. Mol. Sci.} \textbf{2011},
  \emph{1}, 153--163.

\bibitem{science_1991_252_103}
Marder,~S.~R.; Beratan,~D.~N.; Cheng,~L.-T. \emph{Science} \textbf{1991},
  \emph{252}, 103--106.

\bibitem{j_phys_condens_matter_200_15_R897}
Dalton,~L.~R. \emph{J. Phys. Condens. Matter} \textbf{2003}, \emph{15},
  R897--R934.

\bibitem{mol_phys_2003_101_2279}
Squitieri,~E.; Paz,~J.~L.; Hern\'{a}ndez,~A.~J. \emph{Mol. Phys.}
  \textbf{2003}, \emph{101}, 2279--2284.

\bibitem{phys_rev_lett_1999_82_3304}
Wang,~T.; Moll,~N.; Cho,~K.; Joannopoulos,~J.~D. \emph{Phys. Rev. Lett.}
  \textbf{1999}, \emph{82}, 3304--3307.

\bibitem{bera91}
Beratan,~D.~N. In \emph{Materials for Nonlinear Optics: Chemical Perspectives};
  Marder,~S.~R.; Sohn,~J.~E.; Stucky,~G.~D., Eds.;
\newblock American Chemical Society: Washington, DC, \textbf{1991}, pp 89--102.

\bibitem{j_phys_chem_1996_100_10595}
Kuhn,~C.; Beratan,~D.~N. \emph{J. Phys. Chem.} \textbf{1996}, \emph{100},
  10595--10599.

\bibitem{zeitschr_elektrochem_1937_43_752}
H\"{u}ckel,~E. \emph{Zeitschr. Elektrochem. angew. physik. Chem.}
  \textbf{1937}, \emph{43}, 752--788.

\bibitem{zeitschr_elektrochem_1937_43_827}
H\"{u}ckel,~E. \emph{Zeitschr. Elektrochem. angew. physik. Chem.}
  \textbf{1937}, \emph{43}, 827--849.

\bibitem{sci_china_ser_b_2009_52_1769}
Hu,~X.; Beratan,~D.~N.; Yang,~W. \emph{Sci. China Ser. B\,---\,Chem.}
  \textbf{2009}, \emph{52}, 1769--1776.

\bibitem{chem_rev_2012_112_2889}
Le,~T.; Epa,~V.~C.; Burden,~F.~R.; Winkler,~D.~A. \emph{Chem. Rev.}
  \textbf{2012}, \emph{112}, 2889--2919.

\bibitem{curr_med_chem_2011_18_2517}
Mavromoustakos,~T.; Durdagi,~S.; Koukoulitsa,~C.; Simcic,~M.;
  Papadopoulos,~M.~G.; Hodoscek,~M.; Grdadolnik,~S.~G. \emph{Curr. Med. Chem.}
  \textbf{2011}, \emph{18}, 2517--2530.

\bibitem{chem_rev_1996_96_1027}
Karelson,~M.; Lobanov,~V.~S. \emph{Chem. Rev.} \textbf{1996}, \emph{96},
  1027--1043.

\bibitem{chem_soc_rev_1995_24_279}
Katritzky,~A.~R.; Lobanov,~V.~S.; Karelson,~M. \emph{Chem. Soc. Rev.}
  \textbf{1995}, \emph{24}, 279--287.

\bibitem{excli_journal_2009_8_74}
Nantasenamat,~C.; Isarankura-Na-Ayudhya,~C.; Naenna,~T.; Prachayasittikul,~V.
  \emph{EXCLI Journal} \textbf{2009}, \emph{8}, 74--88.

\bibitem{chem_rev_2010_110_5714}
Katritzky,~A.~R.; Kuanar,~M.; Slavov,~S.; Hall,~C.~D.; Karelson,~M.; Kahn,~I.;
  Dobchev,~D.~A. \emph{Chem. Rev.} \textbf{2010}, \emph{110}, 5714--5798.

\bibitem{chempluschem_2012_77_507}
Berhanu,~W.~M.; Pillai,~G.~G.; Oliferenko,~A.~A.; Katritzky,~A.~R.
  \emph{ChemPlusChem} \textbf{2012}, \emph{77}, 507--517.

\bibitem{cros_phd}
Cros,~A. F.~A. Ph.D.\ thesis, University of Strasbourg, 1863.

\bibitem{j_am_chem_soc_1964_86_1616}
Hansch,~C.; Fujita,~T. \emph{J. Am. Chem. Soc.} \textbf{1964}, \emph{86},
  1616--1626.

\bibitem{chem_rev_1935_17_125}
Hammet,~L.~P. \emph{Chem. Rev.} \textbf{1935}, \emph{17}, 125--136.

\bibitem{j_am_chem_soc_1937_59_96}
Hammett,~L.~P. \emph{J. Am. Chem. Soc.} \textbf{1937}, \emph{59}, 96--103.

\bibitem{computdet}
 These data are calculated with the program {\sc Adf}, version
  2010.02b\cite{j_comput_chem_2001_22_931}, utilizing the BP86
  exchange--correlation
  functional\cite{phys_rev_a_1988_38_3098,phys_rev_b_1986_33_8822} and the TZP
  basis set without frozen cores\cite{j_comput_chem_2003_24_1142} at all atoms.
  Scalar-relativistic effects were taken into account by means of the zeroth
  order regular approximation\cite{j_chem_phys_1999_110_8943}.

\bibitem{inorg_chem_2009_48_1638}
Schenk,~S.; Reiher,~M. \emph{Inorg. Chem.} \textbf{2009}, \emph{48},
  1638--1648.

\bibitem{j_chem_inf_comput_sci_1998_38_251}
Zheng,~W.; Cho,~S.~J.; Tropsha,~A. \emph{J. Chem. Inf. Comput. Sci.}
  \textbf{1998}, \emph{38}, 251--258.

\bibitem{j_chem_inf_comput_sci_1998_38_259}
Cho,~S.~J.; Zheng,~W.; Tropsha,~A. \emph{J. Chem. Inf. Comput. Sci.}
  \textbf{1998}, \emph{38}, 259--268.

\bibitem{science_1983_220_671}
Kirkpatrick,~S.; Gelatt,~Jr.,~C.~D.; Vecchi,~M.~P. \emph{Science}
  \textbf{1983}, \emph{220}, 671--680.

\bibitem{parallel_comput_1988_7_65}
M\"uhlenbein,~H.; Gorges-Schleuter,~M.; Kr\"amer,~O. \emph{Parallel Comput.}
  \textbf{1988}, \emph{7}, 65--85.

\bibitem{j_am_stat_assoc_1949_44_335}
Metropolis,~N.; Ulam,~S. \emph{J. Am. Stat. Assoc.} \textbf{1949}, \emph{44},
  335--341.

\bibitem{angew_chem_int_ed_engl_1996_35_1286}
Sch\"on,~J.~C.; Jansen,~M. \emph{Angew. Chem. Int. Ed. Engl.} \textbf{1996},
  \emph{35}, 1286--1304.

\bibitem{angew_chem_int_ed_2002_41_3746}
Jansen,~M. \emph{Angew. Chem. Int. Ed.} \textbf{2002}, \emph{41}, 3746--3766.

\bibitem{ber_bunsenges_phys_chem_1995_99_1148}
Putz,~H.; Sch\"{o}n,~J.~C.; Jansen,~M. \emph{Ber. Bunsenges. Phys. Chem.}
  \textbf{1995}, \emph{99}, 1148--1153.

\bibitem{comput_mater_sci_1995_4_43}
Sch\"{o}n,~J.~C.; Jansen,~M. \emph{Comput. Mater. Sci.} \textbf{1995},
  \emph{4}, 43--58.

\bibitem{phys_rev_lett_1994_72_3851}
Werner,~J.~H.; Kolodinski,~S.; Queisser,~H.~J. \emph{Phys. Rev. Lett.}
  \textbf{1994}, \emph{72}, 3851--3854.

\bibitem{nature_1999_402_60}
Franceschetti,~A.; Zunger,~A. \emph{Nature} \textbf{1999}, \emph{402}, 60--63.

\bibitem{phys_rev_b_1995_51_10795}
Silverman,~A.; Zunger,~A.; Kalish,~R.; Adler,~J. \emph{Phys. Rev. B}
  \textbf{1995}, \emph{51}, 10795--10816.

\bibitem{phys_rev_b_1995_51_17398}
Wang,~L.-W.; Zunger,~A. \emph{Phys. Rev. B} \textbf{1995}, \emph{51},
  17398--17416.

\bibitem{j_chem_phys_1994_100_2394}
Wang,~L.-W.; Zunger,~A. \emph{J. Chem. Phys.} \textbf{1994}, \emph{100},
  2394--2397.

\bibitem{phys_rev_lett_2008_100_186403}
Piquini,~P.; Graf,~P.~A.; Zunger,~A. \emph{Phys. Rev. Lett.} \textbf{2008},
  \emph{100}, 186403.

\bibitem{j_comput_phys_2005_208_735}
Kim,~K.; Graf,~P.~A.; Jones,~W.~B. \emph{J. Comput. Phys.} \textbf{2005},
  \emph{208}, 735--760.

\bibitem{phys_rev_lett_2006_97_046401}
Dudiy,~S.~V.; Zunger,~A. \emph{Phys. Rev. Lett.} \textbf{2006}, \emph{97},
  046401--1.

\bibitem{phys_rev_lett_2002_88_255506}
J\'{o}hannesson,~G.~H.; Bligaard,~T.; Ruban,~A.~V.; Skriver,~H.~L.;
  Jacobsen,~K.~W.; N\o{}rskov,~J.~K. \emph{Phys. Rev. Lett.} \textbf{2002},
  \emph{88}, 255506--1.

\bibitem{nature_chem_2009_1_37}
N\o{}rskov,~J.~K.; Bligaard,~T.; Rossmeisl,~J.; Christensen,~C.~H. \emph{Nature
  Chem.} \textbf{2009}, \emph{1}, 37--46.

\bibitem{j_mater_sci_2012_47_7317}
Hautier,~G.; Jain,~A.; Ong,~S.~P. \emph{J. Mater. Sci.} \textbf{2012},
  \emph{47}, 7317--7340.

\bibitem{mrs_bulletin_2006_31_681}
Marzari,~N. \emph{MRS Bulletin} \textbf{2006}, \emph{31}, 681--687.

\bibitem{mrs_bulletin_2010_35_693}
Ceder,~G. \emph{MRS Bulletin} \textbf{2010}, \emph{35}, 693--701.

\bibitem{phys_rev_lett_2010_105_196403}
Chan,~M. K.~Y.; Ceder,~G. \emph{Phys. Rev. Lett.} \textbf{2010}, \emph{105},
  196403.

\bibitem{appl_phys_lett_mater_2013_1_011022}
Jain,~A.; Ong,~S.~P.; Hautier,~G.; Chen,~W.; Richards,~W.~D.; Dacek,~S.;
  Cholia,~S.; Gunter,~D.; Skinner,~D.; Ceder,~G.; Persson,~K.~A. \emph{Appl.
  Phys. Lett. Mater.} \textbf{2013}, \emph{1}, 011002.

\bibitem{inorg_chem_2011_50_656}
Hautier,~G.; Fischer,~C.; Ehrlacher,~V.; Jain,~A.; Ceder,~G. \emph{Inorg.
  Chem.} \textbf{2011}, \emph{50}, 656--663.

\bibitem{stru13}
Struebing,~H.; Ganase,~Z.; Karamertzanis,~P.~G.; Siougkrou1,~E.; Haycock,~P.;
  Piccione,~P.~M.; Armstrong,~A.; Galindo1,~A.; Adjiman,~C.~S. \emph{Nature
  Chem.} \textbf{2013}, \emph{5}, 952--957.

\bibitem{truh13}
Truhlar,~D.~G. \emph{Nature Chem.} \textbf{2013}, \emph{5}, 902--903.

\bibitem{j_am_chem_soc_2006_128_3228}
Wang,~M.; Hu,~X.; Beratan,~D.~N.; Yang,~W. \emph{J. Am. Chem. Soc.}
  \textbf{2006}, \emph{128}, 3228--3232.

\bibitem{phys_rev_1964_136_864}
Hohenberg,~P.; Kohn,~W. \emph{Phys. Rev.} \textbf{1964}, \emph{136}, 864--871.

\bibitem{phys_rev_lett_2004_92_136404}
Yang,~W.; Ayers,~P.~W.; Wu,~Q. \emph{Phys. Rev. Lett.} \textbf{2004},
  \emph{92}, 146404.

\bibitem{comp_methods_appl_mech_eng_1988_71_197}
Bends\o{}e,~M.~P.; Kikuchi,~N. \emph{Comp. Methods Appl. Mech. Eng.}
  \textbf{1988}, \emph{71}, 197--224.

\bibitem{j_phys_chem_a_2008_112_12203}
Keinan,~S.; Therien,~M.~J.; Beratan,~D.~N.; Yang,~W. \emph{J. Phys. Chem. A}
  \textbf{2008}, \emph{112}, 12203--12207.

\bibitem{j_phys_condens_matter_2007_19_402201}
d'Avezac,~M.; Zunger,~A. \emph{J. Phys. Condens. Matter.} \textbf{2007},
  \emph{19}, 402201.

\bibitem{j_phys_chem_a_2007_111_176}
Keinan,~S.; Hu,~X.; Beratan,~D.~N.; Yang,~W. \emph{J. Phys. Chem. A}
  \textbf{2007}, \emph{111}, 176--181.

\bibitem{j_chem_phys_2008_129_064102}
Hu,~X.; Beratan,~D.~N.; Yang,~W. \emph{J. Chem. Phys.} \textbf{2008},
  \emph{129}, 064102.

\bibitem{j_chem_phys_2008_129_044106}
Xiao,~D.; Yang,~W.; Beratan,~D.~N. \emph{J. Chem. Phys.} \textbf{2008},
  \emph{129}, 044106.

\bibitem{pear84}
Pearl,~J. \emph{{Heuristics: Intelligent Search Strategies for Computer Problem
  Solving}};
\newblock Addison-Wesley: Reading, Massachusetts, \textbf{1984}.

\bibitem{j_chem_phys_2008_129_174105}
Balamurugan,~D.; Yang,~W.; Beratan,~D.~N. \emph{J. Chem. Phys.} \textbf{2008},
  \emph{129}, 174105.

\bibitem{science_2008_321_792}
Cohen,~A.~J.; Mori-S\'{a}nchez,~P.; Yang,~W. \emph{Science} \textbf{2008},
  \emph{321}, 792--794.

\bibitem{int_j_quantum_chem_2013_doi_101002qua24375}
von Lilienfeld,~O.~A. \emph{Int. J. Quantum Chem.} \textbf{2013}, \emph{113},
  1676--1689.

\bibitem{j_chem_phys_2006_125_154104}
von Lilienfeld,~O.~A.; Tuckerman,~M.~E. \emph{J. Chem. Phys.} \textbf{2006},
  \emph{125}, 154104.

\bibitem{j_chem_theory_comput_2007_3_1083}
von Lilienfeld,~O.~A.; Tuckerman,~M.~E. \emph{J. Chem. Theory Comput.}
  \textbf{2007}, \emph{3}, 1083--1090.

\bibitem{j_chem_phys_2009_131_164102}
von Lilienfeld,~O.~A. \emph{J. Chem. Phys.} \textbf{2009}, \emph{131}, 164102.

\bibitem{j_chem_phys_2010_133_084104}
Sheppard,~D.; Henkelman,~G.; von Lilienfeld,~O.~A. \emph{J. Chem. Phys.}
  \textbf{2010}, \emph{133}, 084104.

\bibitem{j_chem_phys_2007_127_064305}
Marcon,~V.; von Lilienfeld,~O.~A.; Andrienko,~D. \emph{J. Chem. Phys.}
  \textbf{2007}, \emph{127}, 064305.

\bibitem{new_j_chem_2007_31_818}
Herrmann,~C.; Neugebauer,~J.; Reiher,~M. \emph{New J. Chem.} \textbf{2007},
  \emph{31}, 818--831.

\bibitem{j_chem_phys_2003_118_1634}
Reiher,~M.; Neugebauer,~J. \emph{J. Chem. Phys.} \textbf{2003}, \emph{118},
  1634--1641.

\bibitem{chimia_2009_63_270}
Kiewisch,~K.; Luber,~S.; Neugebauer,~J.; Reiher,~M. \emph{Chimia}
  \textbf{2009}, \emph{63}, 270--274.

\bibitem{j_chem_phys_2008_129_204103}
Kiewisch,~K.; Neugebauer,~J.; Reiher,~M. \emph{J. Chem. Phys.} \textbf{2008},
  \emph{129}, 204103.

\bibitem{j_chem_phys__2009_130_064105}
Luber,~S.; Neugebauer,~J.; Reiher,~M. \emph{J. Chem. Phys.} \textbf{2009},
  \emph{130}, 064105.

\bibitem{chemphyschem_2009_10_2049}
Luber,~S.; Reiher,~M. \emph{ChemPhysChem} \textbf{2009}, \emph{10}, 2049--2057.

\bibitem{snf}
Neugebauer,~J.; Reiher,~M.; Kind,~C.; Hess,~B.~A. \emph{J. Comput. Chem.}
  \textbf{2002}, \emph{23}, 895.

\bibitem{phys_chem_chem_phys_2004_6_4621}
Reiher,~M.; Neugebauer,~J. \emph{Phys. Chem. Chem. Phys.} \textbf{2004},
  \emph{6}, 4621--4629.

\bibitem{j_comput_chem_2004_25_587}
Neugebauer,~J.; Reiher,~M. \emph{J. Comput. Chem.} \textbf{2004}, \emph{25},
  587--597.

\bibitem{j_phys_chem_a_2004_108_2053}
Neugebauer,~J.; Reiher,~M. \emph{J. Phys. Chem. A} \textbf{2004}, \emph{108},
  2053--2061.

\bibitem{surf_science_2006_600_1891}
Herrmann,~C.; Reiher,~M. \emph{Surf. Science} \textbf{2006}, \emph{600},
  1891--1900.

\bibitem{angew_chem_2006_118_3518}
Adler,~T.~B.; Borho,~N.; Reiher,~M.; Suhm,~M.~A. \emph{Angew. Chem.}
  \textbf{2006}, \emph{118}, 3518--3523.

\bibitem{j_comput_chem_2008_29_2460}
Herrmann,~C.; Neugebauer,~J.; Reiher,~M. \emph{J. Comput. Chem} \textbf{2008},
  \emph{29}, 2460--2470.

\bibitem{j_chem_phys_2010_133_174114}
Kovyrshin,~A.; Neugebauer,~J. \emph{J. Chem. Phys.} \textbf{2010}, \emph{133},
  174114.

\bibitem{chem_phys_2011_391_147}
Kovyrshin,~A.; Neugebauer,~J. \emph{Chem. Phys.} \textbf{2011}, \emph{391},
  147--156.

\bibitem{weym14}
Weymuth,~T.; Reiher,~M. \emph{Int. J. Quantum Chem.} \textbf{2014}, \emph{114}, 838--850.

\bibitem{mrs_proceedings_2013_1524_doi}
Weymuth,~T.; Reiher,~M. \emph{MRS Proceedings} \textbf{2013}, \emph{1524}, DOI:
  10.1557/opl.2012.1764.

\bibitem{sellmann1}
Reiher,~M.; Kirchner,~B.; Hutter,~J.; Sellmann,~D.; Hess,~B.~A. \emph{Chem.
  Eur. J.} \textbf{2004}, \emph{10}, 4443--4453.

\bibitem{haag13}
Haag,~M.~P.; Reiher,~M. \emph{Int. J. Quantum Chem.} \textbf{2013}, \emph{113},
  8--20.

\bibitem{marti2009}
Marti,~K.~H.; Reiher,~M. \emph{J. Comput. Chem.} \textbf{2009}, \emph{30},
  2010--2020.

\bibitem{haag2011}
Haag,~M.~P.; Marti,~K.~H.; Reiher,~M. \emph{ChemPhysChem} \textbf{2011},
  \emph{12}, 3204--3213.

\bibitem{bosson2011}
Bosson,~M.; Grudinin,~S.; Bouju,~X.; Redon,~S. \emph{J. Comput. Phys.}
  \textbf{2011}, \emph{231}, 2581--2598.

\bibitem{bosson2012}
Bosson,~M.; Richard,~C.; Plet,~A.; Grudinin,~S.; Redon,~S. \emph{J. Comput.
  Chem.} \textbf{2012}, \emph{33}, 779--790.

\bibitem{haag2014}
Haag,~M.~P.; Reiher,~M. \emph{Faraday Discuss.} \textbf{2014}, \emph{169}, DOI: 10.1039/C4FD00021H 

\bibitem{j_comput_chem_2001_22_931}
te~Velde,~G.; Bickelhaupt,~F.~M.; van Gisbergen,~S.~J.~A.; Guerra,~C.~F.;
  Baerends,~E.~J.; Snijders,~J.~G.; Ziegler,~T. \emph{J. Comput. Chem.}
  \textbf{2001}, \emph{22}, 931.

\bibitem{phys_rev_a_1988_38_3098}
Becke,~A.~D. \emph{Phys. Rev. A} \textbf{1988}, \emph{38}, 3098.

\bibitem{phys_rev_b_1986_33_8822}
Perdew,~J.~P. \emph{Phys. Rev. B} \textbf{1986}, \emph{33}, 8822.

\bibitem{j_comput_chem_2003_24_1142}
van Lenthe,~E.; Baerends,~E.~J. \emph{J. Comput. Chem.} \textbf{2003},
  \emph{24}, 1142.

\bibitem{j_chem_phys_1999_110_8943}
van Lenthe,~E.; Ehlers,~A.~E.; Baerends,~E.~J. \emph{J. Chem. Phys.}
  \textbf{1999}, \emph{110}, 8943.

\end{thebibliography}
\providecommand{\refin}[1]{\\ \textbf{Referenced in:} #1}

\end{document}